\documentclass[10pt,twoside]{article}
\usepackage{amsmath}
\usepackage{pst-all}
\usepackage{pstcol}
\usepackage{graphicx}
\usepackage{flafter}

\numberwithin{equation}{section}
\newtheorem{theorem}{Theorem}

\newtheorem{example}[theorem]{Example}

\begin{document}

\title{Using symmetries and generating functions to calculate and minimize
moments of inertia}
\author{Rodolfo A. Diaz\thanks{%
radiazs@unal.edu.co}, William J. Herrera\thanks{%
jherreraw@unal.edu.co}, R. Martinez\thanks{%
remartinezm@unal.edu.co} \\
Universidad Nacional de Colombia, \\
Departamento de F\'{\i}sica. Bogot\'{a}, Colombia.}
\date{}
\maketitle

\begin{abstract}
This paper presents some formulae to calculate moments of inertia for solids
of revolution and for solids generated by contour plots. For this, the
symmetry properties and the generating functions of the figures are
utilized. The combined use of generating functions and symmetry properties
greatly simplifies many practical calculations. In addition, the explicit
use of generating functions gives the possibility of writing the moment of
inertia as a functional, which in turn allows to utilize the calculus of
variations to obtain a new insight about some properties of this fundamental
quantity. In particular, minimization of moments of inertia under certain
restrictions is possible by using variational methods.

PACS \{45.40.-F, 46.05.th, 02.30.Wd\}

\end{abstract}

\section{Introduction\label{sec:introd}}

The moment of inertia (MI) plays a fundamental role in the mechanics of the
rigid rotator \cite{mecanica}, and is a useful tool in applied physics and
engineering \cite{engineering}. Hence, its explicit calculation is of
greatest interest. In the literature, there are alternative methods to
facilitate the calculations of MI for beginners \cite{methods}, at a more
advanced level \cite{mathprop}, or using an experimental approach \cite{exp}%
. Most textbooks in mechanics, engineering and calculus show some methods to
calculate MI's for certain types of figures \cite{mecanica, engineering,
calculo}. However, they usually do not exploit the symmetry properties of
the object to make the calculation easier.

In this paper, we show some quite general formulae in which the MI's are
written in terms of generating functions. On one hand, by combining
considerations of symmetry with the generating functions approach, we are
able to calculate the MI's for many solids much easier than using common
techniques. On the other hand, the explicit use of generating functions
permits to express the MI's as functionals; thus, we can use the methods of
the calculus of variations (CV)\ to study the mathematical properties of the
MI for several kind of figures. In particular, minimization processes of
MI's under certain restrictions are developed by applying variational
methods.

The paper is organized as follows: in Secs. \ref{sec:revx}, \ref{sec:revy}
we derive expressions for the MI's of solids of revolution. In Sec. \ref%
{sec:contours} we calculate MI's of solids by using the contour plots of the
figure, finding formulae for thin plates as a special case. Sec. \ref%
{sec:applications}, shows some applications of our formulae, exploring some
properties of the MI by using methods of the CV. Sec. \ref{sec:conclusions},
contains the analysis and conclusions.

\section{MI's for solids of revolution generated around the X-axis\label%
{sec:revx}}

\subsection{Moment of inertia with respect to the X-axis\label{sec:revxX}}

\begin{figure}[tbh]
\begin{center}
\includegraphics[width=6.8cm]{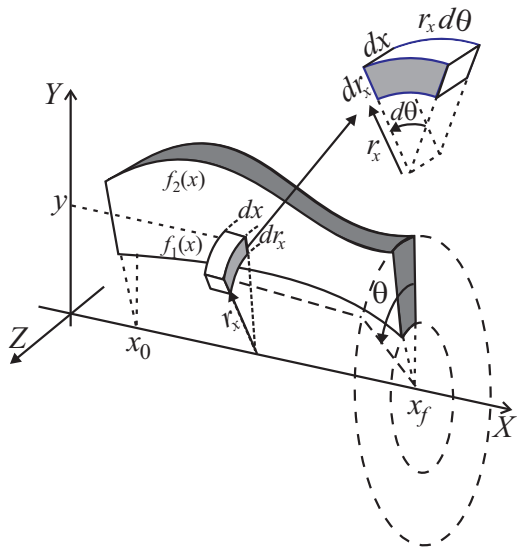}
\end{center}
\caption{\textit{Solid of revolution generated from the $X-$axis. The unit
of integration is a piece of hoop with cross sectional area $\left(
dx\right) \left( dr_{x}\right) $ and length $r_{x}\ d\protect\theta $. We
show a rotated view of the piece of hoop in the upper right corner, for an
easier visualization of its dimensions.}}
\label{fig:revx}
\end{figure}

Let us evaluate the moment of inertia of a solid of revolution with respect
to the axis that generates it, in this case the $X-$axis according to Fig. %
\ref{fig:revx}. We shall assume henceforth, that the solid of revolution is
generated by two functions $f_{1}\left( x\right) $ and $f_{2}\left( x\right) 
$ that fulfill the condition $0\leq $ $f_{1}\left( x\right) \leq f_{2}\left(
x\right) $ for all $x\in \left[ x_{0},x_{f}\right] $.

Owing to the cylindrical symmetry exhibited by the solids of revolution, it
is more convenient to work in a cylindrical system of coordinates. The
coordinates are denoted by $x,r_{x},\theta $ where $r_{x}$ is the distance
from the $X-$axis to the point, the coordinate $\theta $ is defined such
that $\theta =0$ when the vector radius is parallel to the $Y-$axis, and
increases when going from $Y$ (positive) to $Z$ (positive) as shown in Fig. %
\ref{fig:revx}.

We consider a thin hoop with rectangular cross sectional area equal to $%
\left( dx\right) \left( dr_{x}\right) $ and perimeter $2\pi r_{x}$ as Fig. %
\ref{fig:revx} displays. Our infinitesimal element of volume will be a very
short piece of this hoop, lying between $\theta $ and $\theta +d\theta $,
the arc length for this short section of the hoop is $r_{x}\ d\theta $.
Therefore, the infinitesimal element of volume and the corresponding
differential of mass, are given by%
\begin{equation}
dV=\left( dx\right) \left( dr_{x}\right) r_{x}\left( d\theta \right) \ ;\
dm=\rho \left( x,r_{x},\theta \right) \ dV\ ,  \label{revdm}
\end{equation}%
the distance from the $X-$axis$\ $to this element of volume is $r_{x}$.
Therefore, the differential moment of inertia for this element reads%
\begin{equation*}
dI_{X}=r_{x}^{2}\ dm=r_{x}^{3}\rho \left( x,r_{x},\theta \right) \left(
dx\right) \left( dr_{x}\right) \left( d\theta \right) \ ,
\end{equation*}%
and the total moment of inertia is given by%
\begin{equation}
I_{X}=\int_{x_{0}}^{x_{f}}\left\{ \int_{f_{1}\left( x\right) }^{f_{2}\left(
x\right) }\left[ \int_{\theta _{0}}^{\theta _{f}}\rho \left( x,r_{x},\theta
\right) \ d\theta \right] r_{x}^{3}\ dr_{x}\right\} \ dx\ .  \label{revXa}
\end{equation}%
In this expression we first form the complete hoop (no necessarily closed),
it means that we integrate in $\theta $ because when we form the hoop, the $x
$ and $r_{x}$ variables are maintained constant and the integration only
involves $\theta $ as a variable. After the completion of the hoop, we form
a hollow cylinder of minimum radius $f_{1}\left( x\right) $, maximum radius $%
f_{2}\left( x\right) $ and height $dx$. We make it by integrating concentric
hoops, where the radii $r_{x}$\ of the hoops run from $f_{1}\left( x\right) $
to $f_{2}\left( x\right) $. Clearly, the variable $x$ is constant in this
step of the integration. Finally, we integrate the hollow cylinders to
obtain the solid of revolution, it is performed by running the $x$ variable
from $x_{0}\ $to$\ x_{f}$ as we can see in Fig \ref{fig:revx}. This
procedure gives us the formula of Eq. (\ref{revXa}).

The expression given by Eq. (\ref{revXa}) is valid for any solid of
revolution (characterized by the generating functions $f_{1}\left( x\right) $%
, $f_{2}\left( x\right) $) which can be totally inhomogeneous. Even, the
revolution does not have to be from $0$ to $2\pi $. If we assume a complete
revolution for the solid with $\rho =\rho \left( x,r_{x}\right) $ the
formula simplifies to

\begin{equation}
I_{X}=2\pi \int_{x_{0}}^{x_{f}}\left[ \int_{f_{1}\left( x\right)
}^{f_{2}\left( x\right) }\rho \left( x,r_{x}\right) \ r_{x}^{3}\ dr_{x}%
\right] \ dx\ ,  \label{revXb}
\end{equation}%
and even simpler for $\rho =\rho \left( x\right) $%
\begin{equation}
I_{X}=\frac{\pi }{2}\int_{x_{0}}^{x_{f}}\rho \left( x\right) \left[
f_{2}\left( x\right) ^{4}-f_{1}\left( x\right) ^{4}\right] \ dx\ .
\label{revXc}
\end{equation}%
We see that all these expressions for the MI of solids of revolution along
the axis of symmetry, are written in terms of the generating functions $%
f_{1}\left( x\right) $ and $f_{2}\left( x\right) $. In particular, it
follows from Eq. (\ref{revXc}) that for homogeneous solids of revolution
(and even for inhomogeneous ones whose density depend only on $x$ i.e. the
height of the solid), the volume integral involving the calculation of MI is
reduced to a simple integral in one variable. We point out that common
textbooks misuse the cylindrical properties of these type of solids,
evaluating explicitly all the three integrals even for homogeneous objects.

\subsection{Moments of inertia with respect to the $Y$ and $Z\ $axes\label%
{sec:revxYZ}}

\begin{figure}[tbh]
\begin{center}
\includegraphics[width=6.8cm]{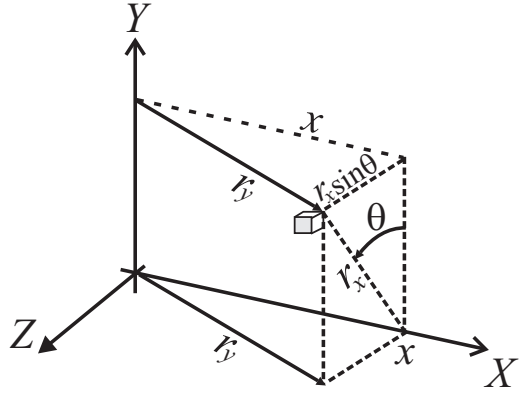}
\end{center}
\caption{\textit{Distance from the $Y-$axis to a differential element of
volume. In terms of the coordinates defined in Sec. \protect\ref{sec:revxX}
the square distance yields $r_{y}^{2}=x^{2}+r_{x}^{2}\sin ^{2}\protect\theta 
$. }}
\label{fig:disty}
\end{figure}

Let us calculate the moment of inertia of the solid of revolution with
respect to the $Y-$axis. We use the same element of volume of the previous
section. The square distance from the $Y-$axis to such element of volume is $%
x^{2}+r_{x}^{2}\sin ^{2}\theta $, as shown in Fig. \ref{fig:disty}.
Therefore, the moment of inertia of this element of volume with respect to $%
Y $ reads%
\begin{equation*}
dI_{Y}=\left( r_{x}^{2}\sin ^{2}\theta +x^{2}\right) \,dm\ ,
\end{equation*}%
with $dm$ given by Eq. (\ref{revdm}). Integrating in a way similar to the
previous section, the MI becomes%
\begin{eqnarray}
I_{Y} &=&\int_{x_{0}}^{x_{f}}\left\{ \int_{f_{1}\left( x\right)
}^{f_{2}\left( x\right) }\left[ \int_{\theta _{0}}^{\theta _{f}}\rho \left(
x,r_{x},\theta \right) \sin ^{2}\theta \ d\theta \right] r_{x}^{3}\
dr_{x}\right\} \ dx+  \notag \\
&&+\int_{x_{0}}^{x_{f}}\left\{ \int_{f_{1}\left( x\right) }^{f_{2}\left(
x\right) }\left[ \int_{\theta _{0}}^{\theta _{f}}\rho \left( x,r_{x},\theta
\right) \ d\theta \right] r_{x}\ dr_{x}\right\} x^{2}\ dx\ .  \label{revYa}
\end{eqnarray}%
Once again, this formula is valid for any inhomogeneous solid of revolution
(even incomplete) generated by the functions $f_{1}\left( x\right) $, $%
f_{2}\left( x\right) $. When assuming a complete solid of revolution with $%
\rho =\rho \left( x,r_{x}\right) $, and taking into account Eq. (\ref{revXb}%
), the latter formula reduces to%
\begin{equation}
I_{Y}=\frac{1}{2}I_{X}+2\pi \int_{x_{0}}^{x_{f}}x^{2}\left[
\int_{f_{1}\left( x\right) }^{f_{2}\left( x\right) }\rho \left(
x,r_{x}\right) \ r_{x}\ dr_{x}\right] \ dx\ .  \label{revYb}
\end{equation}%
From this expression we derive the interesting property $I_{Y}\geq I_{X}/2$,
which is valid for any complete solid of revolution with azimuthal symmetry.
Additionally, for $\rho =\rho \left( x\right) $ we have%
\begin{equation}
I_{Y}=\frac{1}{2}I_{X}+\pi \int_{x_{0}}^{x_{f}}\rho \left( x\right) \ x^{2}%
\left[ f_{2}\left( x\right) ^{2}-f_{1}\left( x\right) ^{2}\right] \ dx\ .
\label{revYc}
\end{equation}

By the same token, for the $Z-$axis, which is perpendicular to the $X$ and $%
Y-$axes and with the origin as the intersection point, we have the following
general formula%
\begin{eqnarray}
I_{Z} &=&\int_{x_{0}}^{x_{f}}\left\{ \int_{f_{1}\left( x\right)
}^{f_{2}\left( x\right) }\left[ \int_{\theta _{0}}^{\theta _{f}}\rho \left(
x,r_{x},\theta \right) \cos ^{2}\theta \ d\theta \right] r_{x}^{3}\
dr_{x}\right\} \ dx+  \notag \\
&&+\int_{x_{0}}^{x_{f}}\left\{ \int_{f_{1}\left( x\right) }^{f_{2}\left(
x\right) }\left[ \int_{\theta _{0}}^{\theta _{f}}\rho \left( x,r_{x},\theta
\right) \ d\theta \right] r_{x}\ dr_{x}\right\} x^{2}\ dx\ .  \label{revZa}
\end{eqnarray}%
We emphasize that in the case of a complete revolution, the formula (\ref%
{revZa}), coincides exactly with $I_{Y}$ in Eq. (\ref{revYb}) when $\rho $
does not depend on $\theta $, as expected from the cylindrical symmetry.
Indeed, for $I_{Z}$ to be equal to $I_{Y}$, the requirement of azimuthal
symmetry could be softened by demanding the conditions 
\begin{equation}
\rho =\rho \left( x,r_{x}\right) \rho \left( \theta \right) \ \ ;\ \
\int_{\theta _{0}}^{\theta _{f}}\rho \left( \theta \right) \sin ^{2}\theta \
d\theta =\int_{\theta _{0}}^{\theta _{f}}\rho \left( \theta \right) \cos
^{2}\theta \ d\theta \ ,  \label{softened}
\end{equation}%
for if the conditions (\ref{softened}) are held, we get that 
\begin{equation*}
\int_{\theta _{0}}^{\theta _{f}}\rho \left( \theta \right) \cos ^{2}\theta \
d\theta =\frac{1}{2}\int_{\theta _{0}}^{\theta _{f}}\rho \left( \theta
\right) \ d\theta \ ,
\end{equation*}%
and $I_{Y}=I_{Z}$ even for an incomplete solid of revolution with no
azimuthal symmetry.

From Eqs. (\ref{revYa}-\ref{revZa}), we see that for the calculation of the
MI for axes perpendicular to the axis of symmetry, we use the same limits of
integration as for the symmetry axis; thus, we do not have to care about the
partitions. Once again, these MI's are written in terms of the generating
functions of the solid $f_{1}\left( x\right) $ and $f_{2}\left( x\right) $.

Finally, we emphasize that textbooks do not usually report the moments of
inertia for solids of revolution with respect to axes perpendicular to the
axis of symmetry. However, they are important in many physical problems. For
instance, a solid of revolution acting as a physical pendulum requires the
calculation of such MI's, see example \ref{bell}.

\begin{figure}[tbh]
\begin{center}
\resizebox{7.0cm}{!}{\mbox{\psset{unit=0.5cm}
\begin{pspicture}(-5,-7)(12,8)
\rput[c]{90}(0,0){
\pspolygon[fillstyle=solid,fillcolor=lightgray,linewidth=0pt](0,-7)(6,-7)(3,0)(1.8,0)(0,-2.5)
\psline[linewidth=1.5pt](1.8,0)(0,-2.5)
\psline[linewidth=1.5pt](-1.8,0)(0,-2.5)
\psellipse[linestyle=dashed,linewidth=1.3pt](0,0)(1.8,0.3)
\psline[linewidth=2.2pt,linecolor=red]{<->}(-1.8,3.5)(0,3.5)
\psline[linewidth=2.2pt,linecolor=blue,linestyle=dashed](-1.8,0)(-1.8,3.5)
\psline[linewidth=1.5pt,linecolor=blue,linestyle=dashed](0,-2.5)(-6,-2.5)
\psline[linewidth=1.5pt,linecolor=blue,linestyle=dashed](-1.8,0)(-6,0)
\psline[linewidth=1.5pt,linecolor=red]{<->}(-6,0)(-6,-2.5)
\psellipse[linestyle=dashed,linewidth=1.3pt](0,0)(3,0.5)
\psellipse[linestyle=dashed,linewidth=1.3pt](0,-7)(6,1)
\psline[linewidth=1.5pt](-6,-7)(-3,0)
\psline[linewidth=1.5pt](6,-7)(3,0)
\psline[linestyle=dashed,linewidth=2.2pt,linecolor=blue](0,0)(0,3.5)
\psline[linestyle=dashed,linewidth=2.2pt,linecolor=blue](3,0)(3,2)
\psline[linecolor=red,linewidth=2.2pt]{<->}(0,2)(3,2)
\psline[linestyle=dashed,linecolor=red,linewidth=2.2pt]{<-}(0,-7)(2.5,-7)
\psline[linestyle=dashed,linecolor=red,linewidth=2.2pt]{->}(3.5,-7)(6,-7)
\psline[linestyle=solid,linecolor=red,linewidth=2.2pt]{<->}(7,-7)(7,0)
\psline[linestyle=dashed,linecolor=green,linewidth=2pt](0,0)(0,-7.8)
\psline[linecolor=green,linewidth=2pt]{->}(0,-8)(0,-11)
\psline[linestyle=dashed,linecolor=green,linewidth=2pt](0,0)(3,0)
\psline[linecolor=green,linewidth=2pt]{->}(3,0)(8,0)
\psline[linecolor=green,linewidth=2pt,linestyle=dashed,dash=3pt 1.5pt](0,0)(-0.75,0.5)
\psline[linecolor=green,linewidth=2pt]{->}(-0.78,0.52)(-6,4)
}
\rput*[c](1.25,-6){$h$}
\uput[r](11,0){\large{$X$}}
\uput[l](0,7.7){\large{$Y$}}
\uput[l](-4,-6){\large{$Z$}}
\rput*[c](-2,1.5){\large{$a_2$}}
\rput[c](7,3.0){\Large{$a_1$}}
\rput*[c](-3.5,-0.9){\footnotesize{$R$}}
\rput*[c](3.5,7){\Large{$H$}}
\uput[u](5,0){$f_{1}\left( x\right)$}
\rput[c]{-30}(1.4,1.4){$f_{1}\left( x\right)$}
\rput[c]{20}(3.5,5){$f_{2}\left( x\right)$}
\end{pspicture}
}}
\end{center}
\caption{\textit{Frustum of a right circular cone with a conical well as a
solid of revolution. The shadowed surface is the one that generates the
solid. $f_{1}\left( x\right) $ and $f_{2}\left( x\right) $ are the functions
that provide the limits of integration.}}
\label{fig:conewell}
\end{figure}
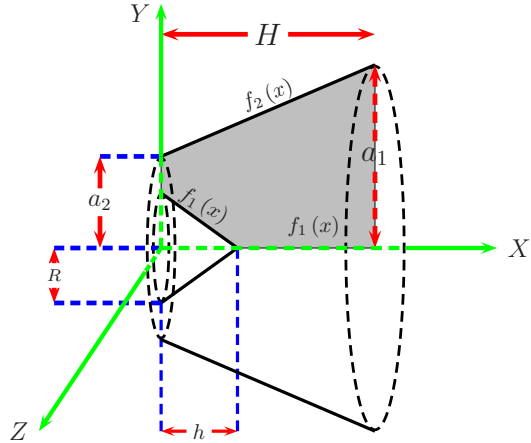

\begin{example}
MI's for a truncated cone with a conical well (see Fig. \ref{fig:conewell}).
The generating functions read%
\begin{eqnarray}
f_{1}\left( x\right) &=&\left\{ 
\begin{array}{ccc}
R\left( 1-\frac{x}{h}\right) & if & x\in \left[ 0,h\right] \\[1mm] 
0 & if & x\in \left( h,H\right]%
\end{array}%
\right.  \notag \\
f_{2}\left( x\right) &=&\left( \frac{a_{1}-a_{2}}{H}\right) x+a_{2}\ ,
\label{funcconewell}
\end{eqnarray}%
where all the dimensions involved are displayed in Fig. \ref{fig:conewell}.
For uniform density, we can replace Eqs. (\ref{funcconewell}) into Eqs. (\ref%
{revXc}, \ref{revYc}) to get%
\begin{eqnarray*}
I_{X} &=&\frac{\pi \rho }{10}\left\{ H\left(
a_{1}^{4}+a_{2}^{4}+a_{1}a_{2}^{3}+a_{1}^{3}a_{2}+\allowbreak
a_{1}^{2}a_{2}^{2}\right) -R^{4}h\right\} \\
I_{Y} &=&I_{Z}=\frac{1}{2}I_{X}+\frac{\pi \rho H^{3}}{5}\left[ \frac{1}{2}%
a_{1}a_{2}+a_{1}^{2}+\frac{1}{6}a_{2}^{2}-\frac{R^{2}}{6}\left( \frac{h}{H}%
\right) ^{3}\right] \ .
\end{eqnarray*}%
It is more usual to give the radius of gyration (RG)\ instead of the MI. For
this we calculate the mass of the solid by using Eq. (\ref{massrevx}),
finding%
\begin{equation}
M=\frac{\pi \rho }{3}\left[ \allowbreak H\left(
a_{1}a_{2}+a_{1}^{2}+a_{2}^{2}\right) -\allowbreak R^{2}h\right] \ .
\label{massconewell}
\end{equation}%
The radii of gyration become%
\begin{eqnarray}
K_{X}^{2} &=&\frac{3\left\{ \allowbreak H\left(
a_{1}^{4}+a_{2}^{4}+a_{1}a_{2}^{3}+a_{1}^{3}a_{2}+\allowbreak
a_{1}^{2}a_{2}^{2}\right) -R^{4}h\right\} }{10\left[ \allowbreak H\left(
a_{1}a_{2}+a_{1}^{2}+a_{2}^{2}\right) -\allowbreak R^{2}h\right] }  \notag \\%
[1mm]
K_{Y}^{2} &=&K_{Z}^{2}=\frac{K_{X}^{2}}{2}+\frac{3}{5}H^{3}\frac{\left[ 
\frac{1}{2}a_{1}a_{2}+a_{1}^{2}+\frac{1}{6}a_{2}^{2}-\frac{R^{2}}{6}\left( 
\frac{h}{H}\right) ^{3}\right] }{\left[ \allowbreak H\left(
a_{1}a_{2}+a_{1}^{2}+a_{2}^{2}\right) -\allowbreak R^{2}h\right] }\ .
\label{Kconewell}
\end{eqnarray}%
By making $R=0$ (and/or $h=0$) we find the RG's for the truncated cone. With 
$R=0$ and $a_{1}=0$, we get the RG's of a cone for which the axes $Y$ and $Z$
pass through its base. Making $R=0$ and $a_{2}=0$, we find the RG's of a
cone but with the axes $Y$ and $Z$ passing through its vertex. Finally, by
setting up $R=0$, and $a_{1}=a_{2}$; we obtain the RG's for a cylinder. In
many cases of interest, we need to calculate the MI's for axes $%
X_{C}Y_{C}Z_{C}$ passing through the center of mass (CM), these MI's can be
calculated by finding the position of the CM with respect to the original
coordinate axes, and using Steiner's theorem (also known as
\textquotedblleft the parallel axis theorem\textquotedblright ). Applying
Eqs. (\ref{CMrevxX}-\ref{CMrevxZ}) the position of the CM for the truncated
cone with a conical well is given by $\left( x_{CM},0,0\right) $ with%
\begin{equation}
x_{CM}=\frac{\left[ \left( 2a_{1}a_{2}+3a_{1}^{2}+a_{2}^{2}\right)
H^{2}-R^{2}h^{2}\right] }{4\left[ \allowbreak H\left(
a_{1}a_{2}+a_{1}^{2}+a_{2}^{2}\right) -\allowbreak R^{2}h\right] }\ .
\label{conewellCM}
\end{equation}%
Gathering Eqs. (\ref{Kconewell},\ \ref{conewellCM}) we find%
\begin{equation*}
K_{X_{C}}^{2}=K_{X}^{2}\ ;\ K_{Y_{C}}^{2}=K_{Y}^{2}-x_{CM}^{2}\ ;\
K_{Z_{C}}^{2}=K_{Z}^{2}-x_{CM}^{2}\ .
\end{equation*}
\end{example}

\subsection{Another alternative of calculation and a proof of consistency 
\label{sec:revconsist}}

In addition to the the parallel and the perpendicular axis theorems, there
is another useful theorem about MI's that is not usually included in common
texts, namely \cite{theorem}%
\begin{equation}
I_{X}+I_{Y}+I_{Z}=2\sum\limits_{i}m_{i}R_{i}^{2}\ ,  \label{theormom}
\end{equation}%
where $X,Y,Z$ are three mutually perpendicular intersecting axes,$\ m_{i}$
is the mass of the $i-$th particle and $R_{i}$ is its distance from the
intersection. We shall see that our general formulae for MI's of solids of
revolution, fulfill the theorem. From Eqs. (\ref{revXa}, \ref{revYa}, \ref%
{revZa}) we have

\begin{equation}
I_{X}+I_{Y}+I_{Z}=2\int_{x_{0}}^{x_{f}}\int_{f_{1}\left( x\right)
}^{f_{2}\left( x\right) }\int_{\theta _{0}}^{\theta _{f}}\left(
x^{2}+r_{x}^{2}\right) \rho \left( x,r_{x},\theta \right) \ r_{x}\ \left(
d\theta \right) \left( dr_{x}\right) \left( dx\right) \ .
\label{provetheor1}
\end{equation}%
Moreover, if we take into account that the distance from the intersecting
point (the origin of coordinates) to the element of volume is $%
R^{2}=x^{2}+r_{x}^{2}$, and using Eq. (\ref{revdm}) we conclude that

\begin{equation}
I_{X}+I_{Y}+I_{Z}=2\int_{V}R^{2}\ dm\ ,  \label{theormom2}
\end{equation}%
which is the continuous version of the theorem established in Eq. (\ref%
{theormom}). As well as providing a proof of consistency, this theorem could
reduce the task to estimate the MI's, especially when a certain spherical
symmetry is involved.

Further, it is interesting to see that the MI's $I_{X},I_{Y},I_{Z},$ in Eqs.
(\ref{revXa},\ \ref{revYa},\ \ref{revZa}) fulfill the triangular inequalities%
\begin{equation*}
I_{X}\leq I_{Y}+I_{Z}\ ,
\end{equation*}%
and same for any cyclic change of the labels. The triangular inequalities
follow directly from the definition of MI, and are valid for any arbitrary
object. Though the demostration of these inequalities is straightforward,
they are not usually considered in the literature. In the case of thin
plates, one of them becomes an equality.

\begin{example}
The following example shows the usefulness of the theorem of Eqs. (\ref%
{theormom}, \ref{theormom2}) in practical calculations. Let us consider the
MI of a sphere centered at the origin, whose density is factorizable in
spherical coordinates such that $\rho =\rho (R)$. Where $R$ is the distance
from the origin of coordinates to the point. The symmetry of $\rho $ leads
to $I_{X}=I_{Y}=I_{Z}$ and the theorem in Eq. (\ref{theormom2}) gives%
\begin{equation}
3I_{X}=8\pi \int_{0}^{R_{0}}\rho \left( R\right) \ R^{4}\ dR\ ,
\label{sphereX1}
\end{equation}%
the mass of the sphere is%
\begin{equation}
M=4\pi \int_{0}^{b}\rho (R)R^{2}\ dR\ ,  \label{spheremass}
\end{equation}%
from which the moment of inertia can be written as%
\begin{equation}
I_{X}=\frac{2}{3}M\frac{\int_{0}^{b}\rho (R)R^{4}\ dR}{\int_{0}^{b}\rho
(R)R^{2}\ dR}\ .  \label{sphereX2}
\end{equation}%
We can calculate for instance, the classical MI of an electron in a
hydrogen-like atom, with respect to an axis that passes through its CM. For
example, for the state $\left( 1,0,0\right) $ we have that $\rho (R)=2\left(
Z/a_{0}\right) ^{3/2}e^{-ZR/a_{0}}$,\ where $Z$ is the atomic number and $%
a_{0}$ the Bohr's radius. The MI becomes%
\begin{equation*}
I_{X}=\frac{8m_{e}}{Z^{2}}a_{0}^{2}\ .
\end{equation*}
\end{example}

\section{MI's of solids of revolution generated around the Y-axis\label%
{sec:revy}}

\begin{figure}[tbh]
\begin{center}
\includegraphics[width=6.8cm]{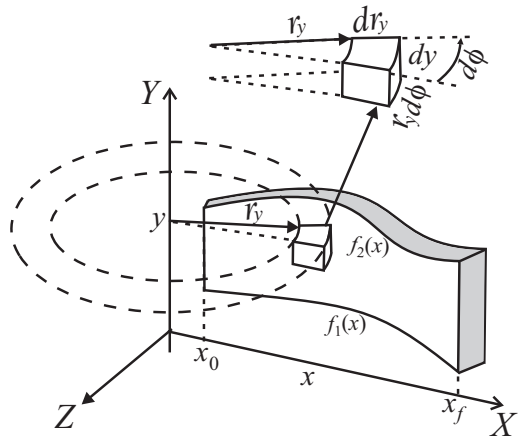}
\end{center}
\caption{\textit{Solid of revolution generated around the $Y-$axis. The unit
of integration is a piece of hoop with cross sectional area $\left(
dy\right) \left( dr_{y}\right) $ and length $r_{y}\ d\protect\phi $.}}
\label{fig:revy}
\end{figure}

By using a couple of generating functions $f_{1}\left( x\right) $ and $%
f_{2}\left( x\right) \ $like in the previous section, we are able to
generate another solid of revolution by rotating such functions around the $%
Y-$axis instead of the $X-$axis as Fig. \ref{fig:revy} displays. In this
case however, we should assume that $x_{0}\geq 0$; such that all points in
the generating surface have always non-negative $x$ coordinates. Instead, we
might allow the functions $f_{1}\left( x\right) $, $f_{2}\left( x\right) $
to be negative though still demanding that $f_{1}\left( x\right) \leq
f_{2}\left( x\right) $ in the whole interval of $x$. In this case, it is
more convenient to use another cylindrical system in which we define the
coordinates $\left( r_{y},y,\phi \right) $, where $r_{y}$ is the distance
from the $Y-$axis to the point, and the angle $\phi $ has been defined such
that $\phi =0$ when the vector radius is parallel to the $Z-$axis
(positive), increasing when going from $Z$ (positive) to $X$ (positive). One
important comment is in order, since the surface that generates the solid
lies on the $XY-$plane, the $x$ coordinate of any point of this surface
(which is always non-negative according to our assumptions) coincides with
the coordinate$\ r_{y}$, therefore we shall write $f_{1}\left( r_{y}\right)
\ $and $f_{2}\left( r_{y}\right) $ instead of $f_{1}\left( x\right) ,\
f_{2}\left( x\right) $ for the functions that bound the generating surface.

The procedure to evaluate the MI in the general case is analogous to the
techniques used in section \ref{sec:revx}, the results are

\begin{eqnarray}
I_{X} &=&\int_{x_{0}}^{x_{f}}\left\{ \int_{f_{1}\left( r_{y}\right)
}^{f_{2}\left( r_{y}\right) }\left[ \int_{\phi _{0}}^{\phi _{f}}\rho \left(
r_{y},y,\phi \right) \cos ^{2}\phi \ d\phi \right] \ dy\right\} r_{y}^{3}\
dr_{y}+  \notag \\
&&+\int_{x_{0}}^{x_{f}}\left\{ \int_{f_{1}\left( r_{y}\right) }^{f_{2}\left(
r_{y}\right) }\left[ \int_{\phi _{0}}^{\phi _{f}}\rho \left( r_{y},y,\phi
\right) d\phi \right] \ y^{2}\ dy\right\} r_{y}\ dr_{y}\ ,  \label{revyXa}
\end{eqnarray}%
\begin{equation}
I_{Y}=\int_{x_{0}}^{x_{f}}\left\{ \int_{f_{1}\left( r_{y}\right)
}^{f_{2}\left( r_{y}\right) }\left[ \int_{\phi _{0}}^{\phi _{f}}\rho \left(
r_{y},y,\phi \right) d\phi \right] \ dy\right\} r_{y}^{3}\ dr_{y}\ ,
\label{revyYa}
\end{equation}%
\begin{eqnarray}
I_{Z} &=&\int_{x_{0}}^{x_{f}}\left\{ \int_{f_{1}\left( r_{y}\right)
}^{f_{2}\left( r_{y}\right) }\left[ \int_{\phi _{0}}^{\phi _{f}}\rho \left(
r_{y},y,\phi \right) \sin ^{2}\phi \,d\phi \right] \ dy\right\} r_{y}^{3}\
dr_{y}+  \notag \\
&&+\int_{x_{0}}^{x_{f}}\left\{ \int_{f_{1}\left( r_{y}\right) }^{f_{2}\left(
r_{y}\right) }\left[ \int_{\phi _{0}}^{\phi _{f}}\rho \left( r_{y},y,\phi
\right) d\phi \right] \ y^{2}\ dy\right\} r_{y}\ dr_{y}\ .  \label{revyZa}
\end{eqnarray}%
As before, these expressions become simpler in the case of a complete
revolution with azimuthal symmetry,%
\begin{equation}
I_{Y}=2\pi \int_{x_{0}}^{x_{f}}\left[ \int_{f_{1}\left( r_{y}\right)
}^{f_{2}\left( r_{y}\right) }\rho \left( r_{y},y\right) \ dy\right] \
r_{y}^{3}\ dr_{y}\ ,  \label{revyYb}
\end{equation}%
\begin{equation}
I_{X}=\frac{1}{2}I_{Y}+2\pi \int_{x_{0}}^{x_{f}}\left[ \int_{f_{1}\left(
r_{y}\right) }^{f_{2}\left( r_{y}\right) }\rho \left( r_{y},y\right) \
y^{2}\ dy\right] r_{y}\ dr_{y}\ ,  \label{revyXb}
\end{equation}%
and in this case $I_{X}=I_{Z}$, as expected from symmetry arguments.
Further, assuming $\rho =\rho \left( r_{y}\right) $ the expressions
simplifies to%
\begin{equation}
I_{Y}=2\pi \int_{x_{0}}^{x_{f}}r_{y}^{3}\ \rho \left( r_{y}\right) \left[
f_{2}\left( r_{y}\right) -f_{1}\left( r_{y}\right) \right] \ dr_{y}\ ,
\label{revyYc}
\end{equation}%
\begin{equation}
I_{X}=I_{Z}=\frac{1}{2}I_{Y}+\frac{2\pi }{3}\int_{x_{0}}^{x_{f}}\rho \left(
r_{y}\right) \left[ \ f_{2}\left( r_{y}\right) ^{3}-f_{1}\left( r_{y}\right)
^{3}\right] r_{y}\ dr_{y}\ .  \label{revyXc}
\end{equation}%
We can verify again that the property $I_{X}=I_{Z}\geq I_{Y}/2$ appears in
the case of azimuthal symmetry, Eq. (\ref{revyXb}). This property is also
satisfied in the case of incomplete solids of revolution, if\ conditions
analogous to Eq. (\ref{softened}) for the $\phi $ angle are fulfilled.
Moreover, the theorem given by Eqs. (\ref{theormom}, \ref{theormom2}) is
also held by these formulae, giving an alternative way for the calculation.
Finally, the triangular inequalities also hold.

These formulae are especially useful in the case in which the generating
functions $f_{1}\left( x\right) $,$\ f_{2}\left( x\right) \ $do not admit
inverses, since in such case we cannot find the corresponding inverse
functions $g_{1}\left( x\right) $,$\ g_{2}\left( x\right) $ to generate the
same figure by rotating around the $X-$axis. This is the case in the
following example

\begin{figure}[tbh]
\psset{unit=0.5cm}
\par
\begin{center}
\resizebox{5.2cm}{!}{\mbox{\begin{pspicture}(-7,-1)(7,5.5)
\pscustom[linestyle=none]{\psline(0,0)(0,2)
\psplot[liftpen=1]{0}{5}{180 x mul sin 2 add}
\psline(5,2)(5,0)
\fill[fillstyle=solid, fillcolor=lightgray]}

\psplot[linewidth=1.5pt]{0}{5}{180 x mul sin 2 add}
\psplot[linewidth=1.5pt,linestyle=dashed]{-5}{0}{180 x mul sin neg 2 add}
\psline[linestyle=dashed](0,0)(0,2)
\psline[linestyle=dashed](5,2)(5,0)
\psline[linestyle=dashed](-5,2)(-5,0)
\psellipse[linewidth=1pt,linestyle=dashed](0,3)(0.5,0.1667)
\psellipse[linewidth=1pt,linestyle=dashed](0,3)(2.5,0.833)
\psellipse[linewidth=1pt,linestyle=dashed](0,3)(4.5,1.5)
\psaxes[ticksize=0.5pt,labels=none]{->}(0,0)(-6,-1)(6,5)
\uput[r](0,5){$Y$}
\uput[r](6,0){$X$}
\uput[d](5,0){\tiny{$(R,0)$}}
\end{pspicture}
}}
\end{center}
\caption{\textit{Solid of revolution created by rotating the generating
functions $f_{1}\left( x\right) =0$, and $f_{2}\left( x\right) =h+A\sin
\left( \frac{n\protect\pi x}{R}\right) $ around the $Y-$axis. The $x\ $%
variable lies in the interval $\left[ 0,R\right] $. From the picture it is
clear that one of the generators do not admit an inverse.}}
\label{fig:sin}
\end{figure}
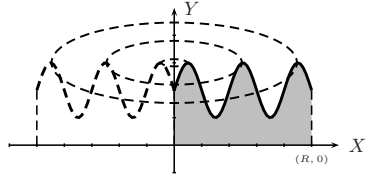

\begin{example}
Calculate the MI's of a homogeneous solid formed by rotating the functions%
\begin{equation}
f_{1}\left( x\right) =0\ ;\ f_{2}\left( x\right) =h+A\sin \left( \frac{n\pi x%
}{R}\right) \ ,  \label{trigfig}
\end{equation}%
around the $Y-axis\ $(see Fig. \ref{fig:sin}), where the functions are
defined in the interval $x\in \left[ 0,R\right] $, and $n$ are positive
integers. We demand $h\geq \left\vert A\right\vert $, if $n>1$; besides, if $%
n=1$ and $\left\vert A\right\vert >h$ we demand $A>0$. These requirements
assure that $f_{2}\left( x\right) \geq f_{1}\left( x\right) $ for all $x\in %
\left[ 0,R\right] $. The mass of the solid, obtained from (\ref{massrevy})
reads%
\begin{equation*}
M=\frac{\pi R^{2}\rho }{n\pi }\left[ n\pi h+2A\left( -1\right) ^{n+1}\right]
\end{equation*}%
and replacing the generating functions into the Eqs. (\ref{revyYc}, \ref%
{revyXc}) we get%
\begin{eqnarray*}
K_{Y}^{2} &=&\frac{R^{2}}{2}\left( \frac{n\pi h+4A(6n^{-2}\pi
^{-2}-1)(-1)^{n}}{n\pi h+2A(-1)^{n+1}}\right) \\[1mm]
K_{X}^{2} &=&K_{Z}^{2}=\frac{K_{Y}^{2}}{2}+\frac{1}{18}\frac{3n\pi
h[2h^{2}+3A^{2}]+4A(-1)^{n+1}[9h^{2}+2A^{2}]}{[n\pi h+2A(-1)^{n+1}]}\ .
\end{eqnarray*}%
the position of the CM (obtained from Eqs. \ref{CMrevyX}-\ref{CMrevyZ}), and
the RG's for axes that pass through the CM read%
\begin{eqnarray*}
\mathbf{r}_{CM} &=&\left( 0,y_{CM},0\right) \ \ ;\ \ y_{CM}=\frac{2n\pi
h^{2}+n\pi A^{2}+8Ah\left( -1\right) ^{n+1}}{4\left[ n\pi h+2A\left(
-1\right) ^{n+1}\right] } \\
K_{X_{C}}^{2} &=&K_{X}^{2}-y_{CM}^{2}\ ;\ K_{Y_{C}}^{2}=K_{Y}^{2}\ ;\
K_{Z_{C}}^{2}=K_{Z}^{2}-y_{CM}^{2}
\end{eqnarray*}%
Observe that $f_{2}\left( x\right) $ does not have inverse. Hence, we cannot
generate the same figure by constructing an equivalent function to be
rotated around the $X-$axis\footnote{%
Indeed, we can find the MI of this solid by rotating around the $X-$axis. We
achieve it by splitting up the figure in several pieces in the $y$
coordinate, such that each interval in $y$ defines a function. However, it
implies to introduce more than two generating functions and the number of
such generators increases with $n$, making the calculation more complex.}.
\end{example}

Finally, it worths pointing out that by considering homogeneous solids of
revolution, we obtain the same expressions derived with a different approach
in Ref. \cite{revus}.

\section{Moments of inertia based on the contour plots of some figures\label%
{sec:contours}}

\begin{figure}[tbh]
\begin{center}
\rotatebox{0}{\includegraphics[width=6.8cm]{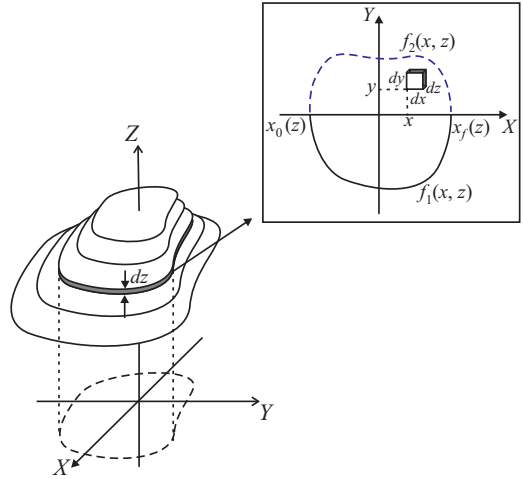}}
\end{center}
\caption{\textit{Contour plots of a solid utilized to calculate its moments
of inertia.}}
\label{fig:contour}
\end{figure}

Suppose that we know the contour plots of certain solid in the $XY$ plane,
i.e. the surfaces shaped by the intersection between planes parallel to the $%
XY$ plane and the solid (see Fig. \ref{fig:contour}). Assume that for a
certain value of the $z$ coordinate, the surface defined by the contour is
bounded by the functions $f_{1}\left( x,z\right) $ and $f_{2}\left(
x,z\right) $ in the $y$ coordinate, and by $x_{0}\left( z\right) $, $%
x_{f}\left( z\right) $ in the $x$ coordinate, as shown in the frame on the
upper right corner of Fig. \ref{fig:contour}. Let us form a thin plate of
thick $dz$ with the surface described above. In turn we can divide such thin
plate into small rectangular boxes with surface $dx\ dy$ and depth $dz$ as
shown in Fig. \ref{fig:contour}, it is well known from the literature \cite%
{mecanica} that the MI with respect to the $X-$axis in cartesian coordinates
reads%
\begin{equation*}
I_{X}=\int_{V}\left( y^{2}+z^{2}\right) \ dm
\end{equation*}%
now, since our infinitesimal elements of volume are rectangular boxes with
volume $dV=dx\ dy\ dz$, the contribution of each rectangular box to the MI
around the $X-$axis is given by%
\begin{equation*}
dI_{X}=\left( y^{2}+z^{2}\right) \ dm=\left( y^{2}+z^{2}\right) \rho \left(
x,y,z\right) \ dx\ dy\ dz\ ,
\end{equation*}%
integrating over all variables, we obtain%
\begin{equation}
I_{X}=\int_{z_{0}}^{z_{f}}\left\{ \int_{x_{0}\left( z\right) }^{x_{f}\left(
z\right) }\left[ \int_{f_{1}\left( x,z\right) }^{f_{2}\left( x,z\right)
}y^{2}\rho \ dy\right] dx\right\} dz+\int_{z_{0}}^{z_{f}}\left\{
\int_{x_{0}\left( z\right) }^{x_{f}\left( z\right) }\left[ \int_{f_{1}\left(
x,z\right) }^{f_{2}\left( x,z\right) }\rho \ dy\right] dx\right\} z^{2}\ dz\
.  \label{contourX}
\end{equation}%
The procedure for $I_{Y}$ and $I_{Z}$ is analogous, the results are.%
\begin{eqnarray}
I_{Y} &=&\int_{z_{0}}^{z_{f}}\left\{ \int_{x_{0}\left( z\right)
}^{x_{f}\left( z\right) }\left[ \int_{f_{1}\left( x,z\right) }^{f_{2}\left(
x,z\right) }\rho \ dy\right] x^{2}\ dx\right\}
dz+\int_{z_{0}}^{z_{f}}\left\{ \int_{x_{0}\left( z\right) }^{x_{f}\left(
z\right) }\left[ \int_{f_{1}\left( x,z\right) }^{f_{2}\left( x,z\right)
}\rho \ dy\right] dx\right\} z^{2}\ dz\ ,  \label{contourY} \\
I_{Z} &=&\int_{z_{0}}^{z_{f}}\left\{ \int_{x_{0}\left( z\right)
}^{x_{f}\left( z\right) }\left[ \int_{f_{1}\left( x,z\right) }^{f_{2}\left(
x,z\right) }y^{2}\rho \ dy\right] dx\right\} dz+\int_{z_{0}}^{z_{f}}\left\{
\int_{x_{0}\left( z\right) }^{x_{f}\left( z\right) }\left[ \int_{f_{1}\left(
x,z\right) }^{f_{2}\left( x,z\right) }\rho \ dy\right] x^{2}\ dx\right\} dz\
.  \label{contourZ}
\end{eqnarray}

Once again, we can check that results (\ref{contourX}, \ref{contourY}, \ref%
{contourZ}) satisfy Eq. (\ref{theormom2}). This equation gives us another
way to calculate the three MI's. Finally, the formulae fulfill the
triangular inequalities.

An interesting case arises when we consider the MI's of thin plates. Suppose
a thin plate lying on the XY plane. $\sigma \left( x,y\right) $ denotes its
surface density. This solid is generated by contour plots with volumetric
density%
\begin{equation}
\rho \left( x,y,z\right) =\sigma \left( x,y\right) \ \delta \left( z\right)
\ ,  \label{Diracdensity}
\end{equation}%
where $\delta \left( z\right) $ denotes Dirac's delta function. Replacing
Eq. (\ref{Diracdensity}) into the general formula (\ref{contourX}) we get%
\begin{eqnarray*}
I_{X} &=&\int_{z_{0}}^{z_{f}}H_{1}\left( z\right) \ \delta \left( z\right) \
dz+\int_{z_{0}}^{z_{f}}H_{2}\left( z\right) \ \delta \left( z\right) \
z^{2}\ dz\ , \\
H_{1}\left( z\right)  &\equiv &\int_{x_{0}\left( z\right) }^{x_{f}\left(
z\right) }\left[ \int_{f_{1}\left( x,z\right) }^{f_{2}\left( x,z\right)
}y^{2}\sigma \left( x,y\right) \ dy\right] dx\ ;\ H_{2}\left( z\right)
\equiv \int_{x_{0}\left( z\right) }^{x_{f}\left( z\right) }\left[
\int_{f_{1}\left( x,z\right) }^{f_{2}\left( x,z\right) }\sigma \left(
x,y\right) \ dy\right] dx\ ,
\end{eqnarray*}%
and using the properties of $\delta \left( z\right) $ we get%
\begin{equation*}
I_{X}=H_{1}\left( 0\right) =\int_{x_{0}\left( 0\right) }^{x_{f}\left(
0\right) }\left[ \int_{f_{1}\left( x,0\right) }^{f_{2}\left( x,0\right)
}y^{2}\sigma \left( x,y\right) \ dy\right] dx\ ,
\end{equation*}%
the $z$ coordinate is evaluated at zero all the time, hence there is only
one contour, we write it simply as%
\begin{equation}
I_{X}=\int_{x_{0}}^{x_{f}}\left[ \int_{f_{1}\left( x\right) }^{f_{2}\left(
x\right) }y^{2}\sigma \left( x,y\right) \ dy\right] dx\ .  \label{thinplateX}
\end{equation}%
Similarly $I_{Y},\ I_{Z}$ can be evaluated replacing (\ref{Diracdensity})
into (\ref{contourY}) and (\ref{contourZ}) 
\begin{equation}
I_{Y}=\int_{x_{0}}^{x_{f}}\left[ \int_{f_{1}\left( x\right) }^{f_{2}\left(
x\right) }\sigma \left( x,y\right) \ dy\right] x^{2}\ dx\ ;\
I_{Z}=I_{X}+I_{Y}\ .  \label{thinplateZ}
\end{equation}%
Hence, Eqs. (\ref{thinplateX}, \ref{thinplateZ}) give us the MI's for a thin
plate delimited by $f_{1}\left( x\right) $ and $f_{2}\left( x\right) $ and
by $x_{0}$,$\ x_{f}$; with surface density $\sigma \left( x,y\right) $. It
worths noting that the second of Eqs. (\ref{thinplateZ}) arises from the
application of (\ref{Diracdensity}) into (\ref{contourZ}) without assuming
the perpendicular axes theorem; showing the consistency of our results%
\footnote{%
For students not accustomed to the Dirac's delta function and its
properties, we basically pass from the volume differential $\rho \ dV$ to
the surface differential $\sigma \ dA$.}. 

The formulae shown in this section are written in terms of generating
functions as in the previous sections. However, these generators are
functions of several variables. In developing these formulae we have not
used any particular symmetry; nevertheless, explicit use of some symmetries
could simplify many specific calculations as shown in the following
examples. 
\begin{figure}[tbh]
\psset{unit=0.5cm}
\par
\begin{center}
\resizebox{7.3cm}{!}{\mbox{\begin{pspicture}(-6,-2)(16,11)
\psellipse(0,7)(3,0.5)
\psellipse(0,0)(6,1)
\psellipse[shadow=true,shadowsize=3pt,shadowangle=-90,gradangle=90,linestyle=dotted,fillstyle=gradient,gradbegin=white,gradend=lightgray,gradmidpoint=0.2](0,4)(4.3,0.717)
\psellipse[shadow=true,shadowsize=3pt,shadowangle=-90,gradangle=90,linestyle=dotted,fillstyle=gradient,gradbegin=white,gradend=lightgray,gradmidpoint=0.2](0,0)(4.3,0.717)
\psline[linestyle=dashed,linecolor=blue,linewidth=1.5pt]{<->}(4.3,0)(4.3,3.8)
\psline[linestyle=dashed,linecolor=blue,linewidth=1.5pt]{<->}(-4.3,0)(-4.3,3.8)
\psellipse[shadow=true,shadowsize=5pt,shadowangle=-45,gradangle=90,linestyle=dotted,fillstyle=gradient,gradbegin=white,gradend=lightgray,gradmidpoint=0.2](10,6)(4.3,2.15)
\psaxes[labels=none,ticksize=0.5pt]{<->}(10,6)(5,3)(15,10)
\psline[linecolor=red,linewidth=1.5pt,linestyle=dashed]{<->}(10,6)(14.3,6)
\psline[linecolor=blue,linewidth=1.5pt,linestyle=dashed]{<->}(10,6)(10,8.15)
\uput[d](12.15,6){$a(z)$}
\uput[l](10,7){$b(z)$}
\pnode(2.5,4.6){A}
\pnode(6,5){B}
\ncarc[linecolor=cyan,linewidth=2pt]{->}{A}{B}
\psline(-6,0)(-3,7)
\psline(6,0)(3,7)
\psline[linestyle=dashed,linecolor=red,linewidth=1.5pt]{<-}(0,7)(1,7)
\psline[linestyle=dashed,linecolor=red,linewidth=1.5pt]{->}(2,7)(3,7)
\psline[linestyle=dashed,linecolor=red,linewidth=1.5pt]{<-}(0,0)(2.5,0)
\psline[linestyle=dashed,linecolor=red,linewidth=1.5pt]{->}(3.5,0)(6,0)
\psline[linestyle=dashed,linecolor=red,linewidth=1.5pt]{<->}(0,0)(0,7)
\psline[linestyle=dashed,linecolor=blue,linewidth=1.5pt]{<-}(0,4)(1.52,4)
\psline[linestyle=dashed,linecolor=blue,linewidth=1.5pt]{->}(2.75,4)(4.3,4)
\uput[r](15,6){$X$}
\uput[r](10,10){$Y$}
\rput*[c](-4.3,1.7){$z$}
\rput*[c](4.3,1.7){$z$}
\rput[c](2.15,4){\footnotesize{$a(z)$}}
\rput[c](3,0){\normalsize{$a_{1}$}}
\rput[c](1.5,7){\normalsize{$a_{2}$}}
\rput*(0,2.5){\footnotesize{$H$}}
\uput[l](0,10){\large{$Z$}}
\uput[d](6.8,3.3){\large{$Y$}}
\uput[r](8,0){\large{$X$}}
\psline[linecolor=green,linewidth=2pt]{->}(6,0)(8,0)
\psline[linecolor=green,linewidth=2pt,linestyle=dashed](0,0)(4.8,2.4)
\psline[linecolor=green,linewidth=2pt]{->}(5.0,2.5)(6.8,3.4)
\psline[linecolor=green,linewidth=2pt]{->}(0,7)(0,10)
\end{pspicture}
}}
\end{center}
\caption{\textit{Frustum of a right elliptical cone. The shadowed surfaces
show a contour for a certain value of the $z$ coordinate, as well as its
projection onto the $XY-$plane.}}
\label{fig:contrunel}
\end{figure}
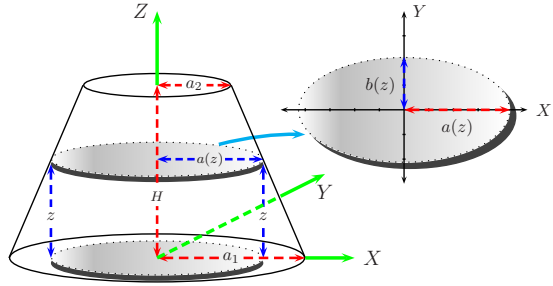

\begin{example}
\label{ex:contrunel}Let us assume a truncated right elliptical cone as shown
in Fig. \ref{fig:contrunel}. Such figure is characterized by the semi-major
and semi-minor axes in the base (denoted by $a_{1}$,$b_{1}$ respectively),
its height $H$, and its semi-major and semi-minor axes in the top (denoted by%
$\ a_{2}$, $b_{2}$ respectively). Suppose that the truncated cone is located
such that the major base lies on the $XY$ plane and the center of such major
base is on the origin of coordinates, as shown in Fig. \ref{fig:contrunel}.
Now, since we are assuming that the figure is not oblique, then all the
contours (see right top on Fig. \ref{fig:contrunel}) are concentric ellipses
centered at the origin, with \emph{the same eccentricity}. Therefore, it is
more convenient to describe such ellipses in terms of their eccentricity $%
\varepsilon \ $and the semi-major axis $a\left( z\right) $. The contours are
then delimited by%
\begin{eqnarray}
f_{1}\left( x,z\right) &=&-\sqrt{\left[ a\left( z\right) ^{2}-x^{2}\right]
\left( 1-\varepsilon ^{2}\right) }\ ;\ f_{2}\left( x,z\right) =\sqrt{\left[
a\left( z\right) ^{2}-x^{2}\right] \left( 1-\varepsilon ^{2}\right) }\ , 
\notag \\[1mm]
x_{0}\left( z\right) &=&-a\left( z\right) \ ;\ x_{f}\left( z\right) =a\left(
z\right) \ ;\ \varepsilon \equiv \sqrt{1-\left( \frac{b\left( z\right) }{%
a\left( z\right) }\right) ^{2}}\ ,  \label{ellipsedelim}
\end{eqnarray}%
where $f_{1,2}\left( x,z\right) $ are the functions that generate the
complete ellipse of semi-major axis $a\left( z\right) $ and eccentricity $%
\varepsilon \ $(independent of $z$). By simple geometric arguments, we could
see that the semi-major axis of one contour of the truncated cone at certain
height $z$ is given by\footnote{%
The semi-minor axes follow a similar equation replacing $a_{1,2}\rightarrow
b_{1,2}$. From such equations we can check that the quotient $\frac{b\left(
z\right) }{a\left( z\right) }$ is constant if we impose $\frac{b_{1}}{a_{1}}=%
\frac{b_{2}}{a_{2}}$. So the latter condition guarantees that the
eccentricity remains constant.}%
\begin{equation}
a\left( z\right) =a_{1}+\left( \frac{a_{2}-a_{1}}{H}\right) z\ ,
\label{radtrunc}
\end{equation}%
Assuming that the density is constant Eq. (\ref{contourZ}) becomes%
\begin{eqnarray*}
I_{Z} &=&\frac{2\rho }{3}\left( \sqrt{1-\varepsilon ^{2}}\right)
^{3}\int_{0}^{H}\left\{ \int_{-a\left( z\right) }^{a\left( z\right) }\left[
\left( \sqrt{a\left( z\right) ^{2}-x^{2}}\right) ^{3}\right] dx\right\} dz \\
&&+2\rho \sqrt{1-\varepsilon ^{2}}\int_{0}^{H}\left\{ \int_{-a\left(
z\right) }^{a\left( z\right) }\left[ \sqrt{a\left( z\right) ^{2}-x^{2}}%
\right] x^{2}\ dx\right\} dz\ ,
\end{eqnarray*}%
where we have already made the integration in $y$. Integration in $x$ yields%
\begin{equation}
I_{Z}=\frac{\pi \rho }{4}\left( 2-\varepsilon ^{2}\right) \sqrt{%
1-\varepsilon ^{2}}\left[ \int_{0}^{H}a\left( z\right) ^{4}\ dz\right] \ ,
\label{contrunnoZ}
\end{equation}%
and taking into account the Eq. (\ref{radtrunc}) we find%
\begin{equation}
I_{Z}=\frac{\pi \rho H}{20}\left( 2-\varepsilon ^{2}\right) \sqrt{%
1-\varepsilon ^{2}}\left[
a_{1}^{4}+a_{2}^{4}+a_{1}a_{2}^{3}+a_{1}^{3}a_{2}+a_{1}^{2}a_{2}^{2}\right]
\ .  \label{contrunnoZ2}
\end{equation}%
Now, the mass of the figure is obtained from Eq. (\ref{masscontour}) and
reads%
\begin{equation}
M=\frac{\pi \rho H}{3}\sqrt{1-\varepsilon ^{2}}\left(
a_{1}^{2}+a_{2}^{2}+a_{1}a_{2}\right) \ ,  \label{masstrunel}
\end{equation}%
therefore, the radius of gyration $K_{Z}^{2}\ $could be written as%
\begin{equation*}
K_{Z}^{2}=\frac{3}{20}\left( 2-\varepsilon ^{2}\right) \left[ \frac{%
a_{1}^{4}+a_{2}^{4}+a_{1}a_{2}^{3}+a_{1}^{3}a_{2}+a_{1}^{2}a_{2}^{2}}{%
a_{1}a_{2}+a_{1}^{2}+a_{2}^{2}}\right] \ .
\end{equation*}%
Further, $K_{X}^{2}$ and $K_{Y}^{2}$ can be derived from Eqs. (\ref{contourX}%
, \ref{contourY}, \ref{masstrunel}) obtaining%
\begin{eqnarray*}
K_{X}^{2} &=&\frac{3}{20}\frac{\left[ \left( 1-\varepsilon ^{2}\right)
\left( a_{1}^{4}+a_{2}^{4}+a_{1}a_{2}^{3}+a_{1}^{3}a_{2}+\allowbreak
a_{1}^{2}a_{2}^{2}\right) +4H^{2}\left( a_{2}^{2}+\frac{1}{6}a_{1}^{2}+\frac{%
1}{2}a_{1}a_{2}\right) \right] }{\left(
a_{1}^{2}+a_{2}^{2}+a_{1}a_{2}\right) } \\[1mm]
K_{Y}^{2} &=&\frac{3}{20}\frac{\left[ \left(
a_{1}^{4}+a_{2}^{4}+a_{1}a_{2}^{3}+a_{1}^{3}a_{2}+\allowbreak
a_{1}^{2}a_{2}^{2}\right) +4H^{2}\left( a_{2}^{2}+\frac{1}{6}a_{1}^{2}+\frac{%
1}{2}a_{1}a_{2}\right) \right] }{\left(
a_{1}^{2}+a_{2}^{2}+a_{1}a_{2}\right) }\ .
\end{eqnarray*}%
when $\varepsilon =0$ we get the radii of gyration of a truncated cone with
circular cross section. In addition, when $\varepsilon =0$ and $a_{2}=0$,
the RG's reduce to the expressions for a cone with the axes $X$ and $Y$
passing through its base. Setting $\varepsilon =0,$ $a_{1}=0$, we also get
the RG's of a cone but with the $X,Y$ axes passing through its vertex. Using 
$\varepsilon =0,$ $a_{1}=a_{2}$ we obtain the RG's of a cylinder. Finally,
with $a_{2}=0$ we get a cone with elliptical cross section, and when $%
a_{1}=a_{2}$ we obtain a cylinder with elliptical cross section. Now, if we
are interested in the MI for coordinates ($X_{C},Y_{C},Z_{C}$) passing
through the CM, we should calculate the position of the CM from Eqs. (\ref%
{CMcontourX}-\ref{CMcontourZ}) and use Steiner's theorem obtaining%
\begin{eqnarray}
\mathbf{r}_{CM} &=&\left( 0,0,z_{CM}\right) \ ;\ z_{CM}=\frac{\left(
2a_{1}a_{2}+a_{1}^{2}+3a_{2}^{2}\right) H}{4\left(
a_{1}a_{2}+a_{1}^{2}+a_{2}^{2}\right) }\ ,  \label{CMtrunel} \\[0.04in]
K_{X_{C}}^{2} &=&K_{X}^{2}-z_{CM}^{2}\ ;\
K_{Y_{C}}^{2}=K_{Y}^{2}-z_{CM}^{2}\ ;\ \ K_{Z_{C}}^{2}=K_{Z}^{2}\ .
\label{KCMtrunel}
\end{eqnarray}
\end{example}

\begin{example}
\label{ex:pyrtrun}Frustum of a right rectangular pyramid: The contour plots
are rectangles. Since the figure is not oblique, the ratios between the
sides of the rectangle are constant. We define $a_{1},b_{1}$ the length and
width of the major base; $a_{2},b_{2}$ the dimensions of the minor base, and 
$H$ the height of the solid, from which we have%
\begin{equation*}
c\equiv \frac{b_{1}}{a_{1}}=\frac{b_{2}}{a_{2}}=\frac{b\left( z\right) }{%
a\left( z\right) }\text{ for all }z\in \left[ 0,H\right] \ ,
\end{equation*}%
where it was assumed that the major base of the figure lies on the $XY$
plane centered in the origin with the lengths $a_{1}$ parallel to the $X-$%
axis and the widths $b_{1}$ parallel to the $Y-$axis. The contours are
delimited by%
\begin{eqnarray*}
f_{1}\left( x,z\right) &=&-\frac{c}{2}a\left( z\right) \ ;\ f_{2}\left(
x,z\right) =\frac{c}{2}a\left( z\right) \ , \\[1mm]
x_{0}\left( z\right) &=&-\frac{a\left( z\right) }{2}\ ;\ x_{f}\left(
z\right) =\frac{a\left( z\right) }{2}\ .
\end{eqnarray*}%
The functional dependence on $z$ is equal to the one in example \ref%
{ex:contrunel}, so $a\left( z\right) \ $is also given by Eq. (\ref{radtrunc}%
). The integration of Eq. (\ref{contourZ}) gives%
\begin{equation*}
I_{Z}=\frac{c\rho }{12}\left( 1+c^{2}\right) \int_{0}^{H}a\left( z\right)
^{4}\ dz\ ,
\end{equation*}%
which is very similar to $I_{Z}$ in Eq. (\ref{contrunnoZ}) for the truncated
cone with elliptical cross section, and since $a\left( z\right) $ in this
example is also given by Eq. (\ref{radtrunc}), the result of $I_{Z}$ for the
truncated pyramid is straightforward by analogy with Eq. (\ref{contrunnoZ2})%
\begin{equation*}
I_{Z}=\frac{c\rho H}{60}\left( 1+c^{2}\right) \left(
a_{1}^{4}+a_{2}^{4}+a_{1}a_{2}^{3}+a_{1}^{3}a_{2}+\allowbreak
a_{1}^{2}a_{2}^{2}\right) \ ,
\end{equation*}%
the mass is obtained from Eq. (\ref{masscontour}) or by analogy with Eq. (%
\ref{masstrunel})%
\begin{equation*}
M=\frac{c\rho H}{3}\left[ a_{1}^{2}+a_{2}^{2}+a_{1}a_{2}\right] \ ,
\end{equation*}%
and the radius of gyration becomes%
\begin{equation*}
K_{Z}^{2}=\frac{\left[ 1+\left( \frac{b_{1}}{a_{1}}\right) ^{2}\right] }{20}%
\left( \frac{a_{1}^{4}+a_{2}^{4}+a_{1}a_{2}^{3}+a_{1}^{3}a_{2}+\allowbreak
a_{1}^{2}a_{2}^{2}}{a_{1}^{2}+a_{2}^{2}+a_{1}a_{2}}\right) \ .
\end{equation*}%
When $a_{2}=0$ we get the RG of a pyramid, if $a_{1}=a_{2}$ we obtain the RG
of the rectangular box. The RG's $K_{X}^{2},$ $K_{Y}^{2}$ are given by%
\begin{eqnarray*}
K_{X}^{2} &=&\frac{3}{5}\frac{\frac{1}{12}\left( \frac{b_{1}}{a_{1}}\right)
^{2}\left( a_{1}^{4}+a_{2}^{4}+a_{1}a_{2}^{3}+a_{1}^{3}a_{2}+\allowbreak
a_{1}^{2}a_{2}^{2}\right) +H^{2}\left( \frac{1}{2}a_{1}a_{2}+\frac{1}{6}%
a_{1}^{2}+a_{2}^{2}\right) }{\left[ a_{1}^{2}+a_{2}^{2}+a_{1}a_{2}\right] }\
, \\[1mm]
K_{Y}^{2} &=&\frac{3}{5}\frac{\frac{1}{12}\left(
a_{1}^{4}+a_{2}^{4}+a_{1}a_{2}^{3}+a_{1}^{3}a_{2}+\allowbreak
a_{1}^{2}a_{2}^{2}\right) +H^{2}\left( \frac{1}{2}a_{1}a_{2}+\frac{1}{6}%
a_{1}^{2}+a_{2}^{2}\right) }{\left[ a_{1}^{2}+a_{2}^{2}+a_{1}a_{2}\right] }\
.
\end{eqnarray*}%
Finally, the expression for the position of the CM coincides with the
results in example \ref{ex:contrunel}, Eq. (\ref{CMtrunel}) with the
corresponding meaning of $a_{1},a_{2}$ in each case. The similarity of all
these results with the ones in example \ref{ex:contrunel}, comes from the
equality in the modulation function of the contours $a\left( z\right) $.
More about it later.
\end{example}

\begin{example}
\label{ex:genellip}\textbf{The general ellipsoid}, centered at the origin of
coordinates is described by%
\begin{equation}
\frac{x^{2}}{a^{2}}+\frac{y^{2}}{b^{2}}+\frac{z^{2}}{c^{2}}=1\ ,
\label{ellipsoid1}
\end{equation}%
we shall assume that $a\geq b\geq c$. A more suitable way to write Eq. (\ref%
{ellipsoid1}) is the following%
\begin{eqnarray}
y^{2} &=&\left( a\left( z\right) ^{2}-x^{2}\right) \left( 1-\varepsilon
^{2}\right) \ ,  \notag \\
a\left( z\right) &\equiv &a\sqrt{1-\frac{z^{2}}{c^{2}}}\text{\ \ };\ b\left(
z\right) \equiv b\sqrt{1-\frac{z^{2}}{c^{2}}}\ ,  \notag \\
\varepsilon &\equiv &\sqrt{1-\left( \frac{b\left( z\right) }{a\left(
z\right) }\right) ^{2}}=\sqrt{1-\left( \frac{b}{a}\right) ^{2}}\ .
\label{ellipsoid3}
\end{eqnarray}%
For fixed values of $z$, we get ellipses whose projections onto the $XY-$%
plane are centered at the origin with semi-major axis $a\left( z\right) $
and semi-minor axis $b\left( z\right) $. The Eqs. (\ref{ellipsoid3}) show
that such ellipses have constant eccentricity, and so we arrive to the
delimited functions of Eqs. (\ref{ellipsedelim}) with $a\left( z\right) $
and $b\left( z\right) $ given by Eqs. (\ref{ellipsoid3}). Therefore, the
first two integrations are performed in the same way as in the truncated
elliptical cone explained in example \ref{ex:contrunel}. Then we can use the
result in Eq. (\ref{contrunnoZ}) (except for the limits of integration in $Z$%
), the last integral is carried out by using Eqs. (\ref{ellipsoid3}).%
\begin{equation*}
I_{Z}=\frac{\pi \rho }{4}\left( 2-\varepsilon ^{2}\right) \sqrt{%
1-\varepsilon ^{2}}\left[ \int_{-c}^{c}a\left( z\right) ^{4}\ dz\right] =%
\frac{4}{15}\pi a^{4}c\rho \left( 2-\varepsilon ^{2}\right) \sqrt{%
1-\varepsilon ^{2}}\ .
\end{equation*}%
The mass of the ellipsoid is given by $M=\left( 4/3\right) \pi \rho
abc=\left( 4/3\right) \pi \rho a^{2}c\sqrt{1-\varepsilon ^{2}}$, so that%
\begin{equation*}
K_{Z}^{2}=\frac{a^{2}\left( 2-\varepsilon ^{2}\right) }{5}=\frac{\left(
a^{2}+b^{2}\right) }{5}\ ,
\end{equation*}%
it could be seen that the RG is independent of $c$, this dependence has been
absorbed into the mass. Similarly, we can get the RG's $K_{X}^{2}$ and $%
K_{Y}^{2}$ applying Eqs. (\ref{contourX}, \ref{contourY}), the results are%
\begin{equation*}
K_{X}^{2}=\frac{1}{5}\left( b^{2}+c^{2}\right) \ ;\ K_{Y}^{2}=\frac{1}{5}%
\left( a^{2}+c^{2}\right) \ .
\end{equation*}%
in this case all axes pass through the center of mass of the object.
\end{example}

Observe that the MI's for the general ellipsoid (example \ref{ex:genellip})
were easily calculated by resorting to the results obtained for the
truncated cone with elliptical cross section (example \ref{ex:contrunel});
it was because both figures have the same type of contours (ellipses) though
in each case such contours are modulated (scaled with the $z$ coordinate) in
different ways. This symmetry between the profiles of both contours
permitted to make the first two integrals in the same way for both figures,
shortening the calculation of the MI for the general ellipsoid considerably.

As for the truncated cone with elliptical cross section (example \ref%
{ex:contrunel}) and the truncated rectangular pyramid (example \ref%
{ex:pyrtrun}), they show the opposite case, i. e. they have different
contours but the modulation is of the same type, this symmetry also
facilitates the calculation of the MI of the truncated pyramid. We emphasize
that this kind of symmetries in either the contours or modulations, can be
exploited for a great variety of figures to simplify the calculation of
their MI's.

\section{Applications utilizing the calculus of variations\label%
{sec:applications}}

\begin{figure}[tbh]
\begin{center}
\rotatebox{0}{\includegraphics[width=3.5cm]{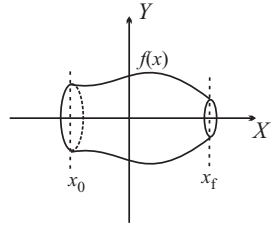}}
\end{center}
\caption{\textit{Optimization of the generating function to minimize the
moment of inertia of a solid of revolution, the mass is the constraint and
the solid lies in the interval $\left[ x_{0},x_{f}\right] $ of length $L$.}}
\label{fig:variac1}
\end{figure}

In all the equations shown in this paper, the MI's can be seen as
functionals of some generating functions. For simplicity, we take a
homogeneous solid of complete revolution around the $X-$axis with $%
f_{1}\left( x\right) =0$. The MI's are functionals of the remaining
generating function, from Eqs. (\ref{revXc}, \ref{revYc}) and relabeling $%
f_{2}\left( x\right) \equiv f\left( x\right) $, we get{\small \ 
\begin{eqnarray}
I_{X}[f] &=&\frac{\pi \rho }{2}\int_{x_{0}}^{x_{f}}f\left( x^{\prime
}\right) ^{4}\ dx^{\prime }\ ,  \label{Ix[f]} \\
I_{Y}[f] &=&\frac{I_{X}[f]}{2}+\pi \rho \int_{x_{0}}^{x_{f}}x^{\prime
2}f\left( x^{\prime }\right) ^{2}\ dx^{\prime }\ .  \label{Iy[f]}
\end{eqnarray}%
}

Then, we can use the methods of the calculus of variations (CV) \footnote{%
{\small The reader not familiarized with the methods of the CV, could skip
this section without sacrifying the understanding of the rest of the
content. Interested readers can look up in the extensive bibliography
concerning this topic, e.g. Ref. \cite{variac1}.}}, in order to optimize the
MI. To figure out possible applications, imagine that we should build up a
figure such that under certain restrictions (that depend on the details of
the design) we require a minimum of energy to set the solid at certain
angular velocity starting from rest. Thus, the optimal design requires the
moment of inertia around the axis of rotation to be a minimum.

As an specific example, suppose that we have a certain amount of material
and we wish to make up a solid of revolution of a fixed length with it, such
that its MI around a certain axis becomes a minimum. To do it, let us
consider a fixed interval $\left[ x_{0},x_{f}\right] $ of length $L$, to
generate a solid of revolution of mass $M$ and constant density $\rho $ (see
Fig. \ref{fig:variac1}). Let us find the function $f\left( x\right) $, such
that $I_{X}$ or $I_{Y}$ become a minimum. Since the mass is kept constant,
we use it as the fundamental constraint{\small \vspace{0mm} 
\begin{equation}
M=\pi \rho \int_{x_{0}}^{x_{f}}f\left( x^{\prime }\right) ^{2}\ dx^{\prime }=%
\text{constant.}  \label{masslig}
\end{equation}%
}

In order to minimize $I_{X}$ we should minimize the functional{\small 
\vspace{0mm} 
\begin{equation}
G_{X}\left[ f\right] =\int_{x_{0}}^{x_{f}}g\left( f,x^{\prime }\right) \
dx^{\prime }=I_{X}\left[ f\right] -\lambda \pi \rho
\int_{x_{0}}^{x_{f}}f\left( x^{\prime }\right) ^{2}\ dx^{\prime }\ \ 
\label{GX[f]}
\end{equation}%
}where $\lambda $ is the Lagrange's multiplicator associated with the
constraint (\ref{masslig}). In order to minimize $G_{X}\left[ f\right] $,\
we should use the Euler-Lagrange equation \cite{variac1}{\small \ 
\begin{equation}
\frac{\delta G_{X}[f]}{\delta f(x)}=\frac{\partial g(f,x)}{\partial f}-\frac{%
\partial }{\partial x}\frac{\partial g(f,x)}{\partial (df/dx)}=0\ \ 
\label{CV}
\end{equation}%
}obtaining{\small \ 
\begin{equation}
\frac{\delta G_{X}[f]}{\delta f(x)}=2\pi \rho f\left( x\right) ^{3}-2\pi
\lambda \rho f\left( x\right) =0\ ,
\end{equation}%
}whose non-trivial solution is given by{\small \ 
\begin{equation}
f\left( x\right) =\sqrt{\lambda }\equiv R\ .
\end{equation}%
}

Analizing the second variational derivative we realize that this condition
corresponds to a minimum. Hence, $I_{X}$ becomes minimum under the
assumptions above for a cylinder of radius $\sqrt{\lambda }$, such radius
can be obtained from the condition (\ref{masslig}), yielding $R^{2}=M/\pi
\rho L$ and $I_{X}$ becomes{\small \ 
\begin{equation}
I_{X,cylinder}=\frac{1}{2}\frac{M^{2}}{\pi \rho L}\ .
\end{equation}%
}

Now, we look for a function that minimizes the MI of the solid of revolution
around an axis perpendicular to the axis of symmetry. From Eqs. (\ref{Iy[f]}%
, \ref{masslig}), we see that the functional to minimize is\vspace{-2mm}%
{\small \ 
\begin{equation}
G_{Y}[f]=\frac{I_{X}[f]}{2}+\pi \rho \int_{x_{0}}^{x_{f}}x^{\prime 2}f\left(
x^{\prime }\right) ^{2}\ dx^{\prime }-\lambda \pi \rho
\int_{x_{0}}^{x_{f}}f\left( x^{\prime }\right) ^{2}\ dx^{\prime }\ ,
\label{GY[f]}
\end{equation}%
}making the variation of $G_{Y}[f]$ with respect to $f\left( x\right) $ we
get

{\small 
\begin{equation}
f\left( x\right) ^{2}=2\left( \lambda -x^{2}\right) \equiv R^{2}-2x^{2}\ ,
\label{fcirculo}
\end{equation}%
}where we have written $2$$\lambda =R^{2}$. By taking $x_{0}=-L/2,\
x_{f}=L/2 $, the function obtained is an ellipse centered at the origin with
semimajor axis $R\ $along the $Y-$axis$,$ semiminor axis $R/\sqrt{2}\ $along
the $X-$axis, and with eccentricity $\varepsilon =1/\sqrt{2}$. When it is
revolved we get an ellipsoid of revolution (spheroid); such spheroid is the
solid of revolution that minimizes the MI with respect to an axis
perpendicular to the axis of revolution. From the condition (\ref{masslig})
we find{\small \ 
\begin{equation}
R^{2}=\frac{M}{\pi \rho L}+\frac{L^{2}}{6}\ .  \label{radioesfera}
\end{equation}%
\begin{figure}[tbh]
\begin{center}
{\small \rotatebox{0}{\includegraphics[width=5.0cm]{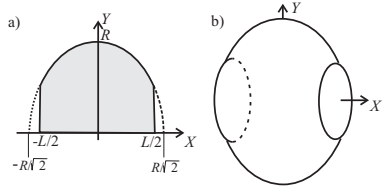}}  }
\end{center}
\caption{(a)\ \textit{Elliptical function that generates the solids of
revolution that minimize$\ I_{Y}$.\ The shaded region is the one that
generates the solid. (b) Truncated spheroid obtained when the shaded region
is revolved.}}
\label{fig:variac2}
\end{figure}
}

In the most general case, the spheroid generated this way is truncated, as
it is shown in Fig. \ref{fig:variac2}, and the condition $R\geq L/\sqrt{{2}}$%
\ should be fulfilled for $f\left( x\right) $\ to be real. The spheroid is
complete when $R=L/\sqrt{{2}}$, and the mass obtained in this case is the
minimum one for the spheroid to fill up the interval $[-L/2,L/2]$, this
minimum mass is given by{\small \vspace{-3mm} 
\begin{equation}
M_{\min }=\frac{\pi \rho L^{3}}{3}\ ,  \label{Mmin}
\end{equation}%
}from (\ref{Iy[f]}), (\ref{fcirculo}), (\ref{radioesfera}) and (\ref{Mmin})
we find{\small \ 
\begin{equation}
I_{Y,spheroid}=\frac{M_{\min }L^{2}\left[ 5\mu +5\mu ^{2}-1\right] }{60}\ ;\
\mu \equiv \frac{M}{M_{\min }}.
\end{equation}%
}Assuming that the densities and masses of the spheroid and the cylinder
coincide, we estimate the quotients{\small \ 
\begin{align}
\frac{I_{Y,cylinder}}{I_{Y,spheroid}}& =\frac{\left( 5\mu +5\mu
^{2}-1\right) }{5\mu \left( \mu +1\right) }<1,  \notag \\
\frac{I_{X,cylinder}}{I_{X,spheroid}}& =\left( 1+\frac{1}{5}\mu ^{-2}\right)
^{-1}<1\ \ .  \label{quotients}
\end{align}%
}

Eqs.\ (\ref{quotients}) show that $I_{Y,sph}<I_{Y,cyl}$ while $%
I_{X,cyl}<I_{X,sph}$. In both cases if $M>>M_{\min }$ the MI's of the
spheroid and the cylinder coincide, it is because the truncated spheroid
approaches the form of a cylinder when the amount of mass to be distributed
in the interval of length $L$ is increased.

On the other hand, in many applications what really matters are the MI's
around axes passing through the CM. In the case of homogeneous solids of
revolution the axis that generates the solid passes through the CM, but this
is not necessarily the case for an axis perpendicular to the former. If we
are interested in minimizing $I_{Y_{C}}$, i.e. the MI with respect to an
axis parallel to $Y$ and passing through the CM, we should write the
expression for $I_{Y_{C}}\ $by\ using the parallel axis theorem and by
combining Eqs. (\ref{Iy[f]}, \ref{masslig}, \ref{CMrevxX}){\small \ 
\begin{align}
I_{Y_{C}}[f]& =\frac{I_{X}[f]}{2}+\pi \rho \int_{x_{0}}^{x_{f}}x^{\prime
2}f\left( x^{\prime }\right) ^{2}\ dx^{\prime }  \notag \\
& -\frac{\pi \rho }{\int_{x_{0}}^{x_{f}}f\left( x^{\prime }\right)
^{2}dx^{\prime }}\left[ \int_{x_{0}}^{x_{f}}x^{\prime }f(x^{\prime })^{2}\
dx^{\prime }\right] ^{2}\ ,
\end{align}%
}thus, the functional to be minimized is{\small \ 
\begin{equation}
G_{Y_{C}}\left[ f\right] =I_{Y_{C}}[f]-\lambda \pi \rho
\int_{x_{0}}^{x_{f}}f\left( x^{\prime }\right) ^{2}\ dx^{\prime }\ ,
\end{equation}%
}after some algebra, we arrive to the following minimizing function{\small \ 
\begin{equation}
f\left( x\right) ^{2}=R^{2}-2\left( x-x_{CM}\right) ^{2}\ ,
\end{equation}%
}where we have written $2\lambda =R^{2}$. It corresponds to a spheroid
(truncated, in general) centered at the point $\left( x_{CM},0,0\right) $ as
expected, showing the consistency of the method.

Finally, it worths remarking that the techniques of the CV shown here can be
extrapolated to more complex situations, as long as we are able to see the
MI's as functionals of certain generating functions. The situations shown
here are simple for pedagogical reasons, but they open a window for other
applications with other constraints\footnote{{\small Another possible
strategy consists of parameterizing the function $f\left( x\right) $, and
find the optimal values for the parameters.}},$\ $for which the minimization
cannot be done by intuition. For instance, if our constraint consists of
keeping the surface constant, the solutions are not easy to guess, and
should be developed by variational methods.

\section{Analysis and conclusions\label{sec:conclusions}}

Most textbooks report MI's for only a few number of simple figures. By
contrast, the examples illustrated in this work have been chosen to be more
general, and can also cover many particular cases. On the other hand, in the
specific case of solids of revolution, only the MI with respect to the
symmetry axis is usually reported. Perhaps the most advantageous feature of
the methods developed in this paper is that the three moments of inertia $%
I_{X},I_{Y},I_{Z}$\ can be calculated by applying the same limits of
integration, and we do not have to worry about the partitions. It is because
all three moments of inertia are written in terms of the generating
functions of the solid. For instance, any solid of revolution acting as a
physical pendulum provides an example in which the MI with respect to an
axis perpendicular to the axis of symmetry is necessary, an specific case is
example \ref{bell} for the Gaussian bell (see appendix \ref{ap:revexamp}).
Moreover, we examine the conditions for the perpendicular MI's to be
degenerate, we find that this degeneracy occurs even for inhomogeneous
solids as long as the density has an azimuthal symmetry. Remarkably, even
for incomplete solids of revolution with the azimuthal symmetry broken, such
degeneracy may occur under certain conditions.

Finally, we point out that for solids of revolution in which densities
depend only on the height of the solid, the expressions for the MI become
simple integrals, such fact makes the integration process much easier.
Simple integrals are advantageous even in the case in which we cannot
evaluate them analytically. Numerical methods to evaluate MI's utilize
typically the geometrical shape of the body; instead, numerical methods for
simple integrals are usually easier to manage.

As for the technique of contour plots, we can realize that many different
figures could have the same type of contours though a different modulation
of them, one specific example is the case of a cone with elliptical cross
section and a general ellipsoid, in both solids the contours are ellipses
but they are modulated (scaled with the $z$ coordinate) in different ways.
On the other hand, in some cases the contours are different but the
modulation is of the same type, for example a cone and a pyramid has the
same type of modulation (scaling) but their contours are totally different.
In both situations we can save a lot of effort by making profit from the
similarities. The reader can check the examples \ref{ex:contrunel}, \ref%
{ex:pyrtrun} and \ref{ex:genellip}, in order to figure out the way in which
we can exploit these symmetries in practical calculations.

Furthermore, textbooks always consider homogeneous figures to estimate the
MI, assumption that is not always realistic. As for our formulae, though
they simplify considerably when we consider homogeneous bodies, the methods
are tractable in many cases when inhomogeneous objects are considered,
allowing more realistic results.

Another interesting remark is that these methods can be used to calculate
products of inertia in the case in which the complete tensor of inertia is
necessary. Moreover, expressions for the CM proceed by similar arguments,
and the appendix \ref{ap:CM}, shows some formulae for the CM of solids of
revolution and solids built up by contour plots. In such expressions we
realize that the same limits of integration defined for the calculation of
the MI's are used to calculate the CM.

Finally, since all our formulae for MI's depend on certain generating
functions, we can see the MI of a wide variety of figures as a functional,
making the MI's suitable to utilize methods of the calculus of variations.
In particular, minimization of the MI under certain restrictions is possible
utilizing variational methods, it could be very useful in applied physics
and engineering.

The authors acknowledge to Dr. H\'{e}ctor M\'{u}nera for revising the
manuscript.

\setcounter{equation}{0}\appendix

\section{Calculation of centers of mass\label{ap:CM}}

\subsection{CM of solids of revolution generated around the $X-axis$\label%
{ap:revxCM}}

Taking into account the definition of the center of mass for continuous
systems%
\begin{equation*}
\vec{r}_{CM}=\frac{\int \vec{r}\ dm}{\int dm}\ ,
\end{equation*}%
we can get general formulae to calculate the CM of a solid by using a
similar procedure to the one followed to get MI's. First of all we calculate
the total mass based on the density and the geometrical shape of the body.
In the case of solids of revolution around the $X-$axis, the total mass of
the solid is given by%
\begin{equation}
M=\int dm=\int_{x_{0}}^{x_{f}}\left\{ \int_{f_{1}\left( x\right)
}^{f_{2}\left( x\right) }\left[ \int_{\theta _{0}}^{\theta _{f}}\rho \left(
x,r_{x},\theta \right) \ d\theta \right] r_{x}\ dr_{x}\right\} \ dx\ ,
\label{massrevx}
\end{equation}%
and the CM coordinates read%
\begin{eqnarray}
X_{CM} &=&\frac{1}{M}\int_{x_{0}}^{x_{f}}\left\{ \int_{f_{1}\left( x\right)
}^{f_{2}\left( x\right) }\left[ \int_{\theta _{0}}^{\theta _{f}}\rho \left(
x,r_{x},\theta \right) \ d\theta \right] r_{x}\ dr_{x}\right\} x\ dx\ ,
\label{CMrevxX} \\
Y_{CM} &=&\frac{1}{M}\int_{x_{0}}^{x_{f}}\left\{ \int_{f_{1}\left( x\right)
}^{f_{2}\left( x\right) }\left[ \int_{\theta _{0}}^{\theta _{f}}\rho \left(
x,r_{x},\theta \right) \cos \theta \ d\theta \right] r_{x}^{2}\
dr_{x}\right\} \ dx\ ,  \label{CMrevxY} \\
Z_{CM} &=&\frac{1}{M}\int_{x_{0}}^{x_{f}}\left\{ \int_{f_{1}\left( x\right)
}^{f_{2}\left( x\right) }\left[ \int_{\theta _{0}}^{\theta _{f}}\rho \left(
x,r_{x},\theta \right) \sin \theta \ d\theta \right] r_{x}^{2}\
dr_{x}\right\} \ dx\ ,  \label{CMrevxZ}
\end{eqnarray}%
the limits of integrations are the ones defined in Sec. \ref{sec:revxX}. In
the case of complete revolution with $\rho =\rho \left( x,r_{x}\right) $, we
obtain $Y_{CM}=Z_{CM}=0$, as expected from the cylindrical symmetry.

\subsection{CM of solids of revolution generated around the $Y-axis$\label%
{ap:revyCM}}

By using the coordinate system and the limits of integration defined in Sec. %
\ref{sec:revy}, we can evaluate the CM for solids of revolution generated
around the $Y-$axis obtaining

\begin{eqnarray}
X_{CM} &=&\frac{1}{M}\int_{x_{0}}^{x_{f}}\left\{ \int_{f_{1}\left(
r_{y}\right) }^{f_{2}\left( r_{y}\right) }\left[ \int_{\phi _{0}}^{\phi
_{f}}\sin (\phi )\rho \left( r_{y},y,\phi \right) \ d\phi \right] \
dy\right\} r_{y}^{2}\ dr_{y}\ ,  \label{CMrevyX} \\
Y_{CM} &=&\frac{1}{M}\int_{x_{0}}^{x_{f}}\left\{ \int_{f_{1}\left(
r_{y}\right) }^{f_{2}\left( r_{y}\right) }\left[ \int_{\phi _{0}}^{\phi
_{f}}\rho \left( r_{y},y,\phi \right) \ d\phi \right] \ y\ dy\right\} r_{y}\
dr_{y}\ ,  \label{CMrevyY} \\
Z_{CM} &=&\frac{1}{M}\int_{x_{0}}^{x_{f}}\left\{ \int_{f_{1}\left(
r_{y}\right) }^{f_{2}\left( r_{y}\right) }\left[ \int_{\phi _{0}}^{\phi
_{f}}\rho \left( r_{y},y,\phi \right) \cos (\phi )\ d\phi \right] \
dy\right\} r_{y}^{2}\ dr_{y}\ ,  \label{CMrevyZ} \\
M &=&\int_{x_{0}}^{x_{f}}\left\{ \int_{f_{1}\left( r_{y}\right)
}^{f_{2}\left( r_{y}\right) }\left[ \int_{\phi _{0}}^{\phi _{f}}\rho \left(
r_{y},y,\phi \right) \ d\phi \right] \ dy\right\} r_{y}\ dr_{y}\ .
\label{massrevy}
\end{eqnarray}

In the case of a complete revolution with $\rho =\rho \left( r_{y},y\right) $%
, we get $X_{CM}=Z_{CM}=0$, due to the cylindrical symmetry.

\subsection{CM of solids formed by contour plots\label{ap:contoursCM}}

In this case, we use the limits of integration defined in Sec. \ref%
{sec:contours}. The CM coordinates read%
\begin{eqnarray}
x_{CM} &=&\frac{1}{M}\int_{z_{0}}^{z_{f}}\left\{ \int_{x_{0}\left( z\right)
}^{x_{f}\left( z\right) }\left[ \int_{f_{1}\left( x,z\right) }^{f_{2}\left(
x,z\right) }\rho \left( x,y,z\right) \ dy\right] x\ dx\right\} dz\ ,
\label{CMcontourX} \\
y_{CM} &=&\frac{1}{M}\int_{z_{0}}^{z_{f}}\left\{ \int_{x_{0}\left( z\right)
}^{x_{f}\left( z\right) }\left[ \int_{f_{1}\left( x,z\right) }^{f_{2}\left(
x,z\right) }\rho \left( x,y,z\right) \ y\ dy\right] \ dx\right\} dz\ ,
\label{CMcontourY} \\
z_{CM} &=&\frac{1}{M}\int_{z_{0}}^{z_{f}}\left\{ \int_{x_{0}\left( z\right)
}^{x_{f}\left( z\right) }\left[ \int_{f_{1}\left( x,z\right) }^{f_{2}\left(
x,z\right) }\rho \left( x,y,z\right) \ dy\right] \ dx\right\} z\ dz\ ,
\label{CMcontourZ} \\
M &=&\int_{z_{0}}^{z_{f}}\left\{ \int_{x_{0}\left( z\right) }^{x_{f}\left(
z\right) }\left[ \int_{f_{1}\left( x,z\right) }^{f_{2}\left( x,z\right)
}\rho \left( x,y,z\right) \ dy\right] \ dx\right\} dz\ .  \label{masscontour}
\end{eqnarray}

\section{Some additional examples for the calculation of moments of inertia 
\label{ap:examples}}

In this appendix we carry out additional calculations of moments of inertia
for some specific figures, by applying the formulae written in sections \ref%
{sec:revx}, \ref{sec:revy} and \ref{sec:contours}; in order to illustrate
the power of the methods.

\subsection{Examples of moments of inertia for solids of revolution\label%
{ap:revexamp}}

\begin{figure}[tbh]
\begin{center}
\includegraphics[width=9.5cm]{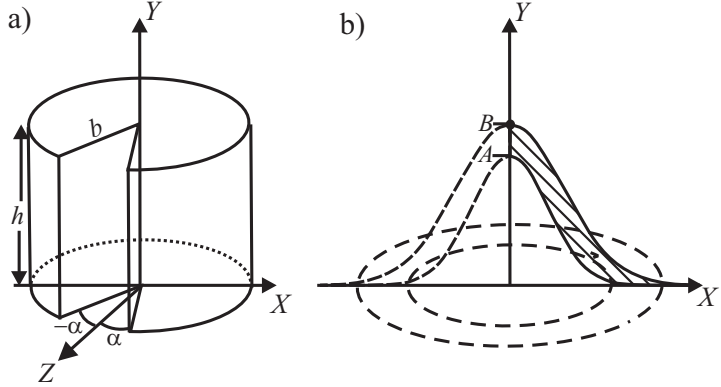}
\end{center}
\caption{\textit{On left: cylindrical wedge generated around the $Y-$axis.
On right: A bell modelated by two Gauss' distributions rotating around the $%
Y-$axis, the bell tolls around an axis perpendicular to the sheet that
passes through the point $B$}}
\label{fig:exrevy}
\end{figure}

\begin{example}
MI for a cylindrical wedge (see Fig. \ref{fig:exrevy}). Let us consider a
cylinder of height $h$ and radius $b$, whose density is given by%
\begin{equation*}
\rho (\phi )=\left\{ 
\begin{array}{c}
\rho \text{, \ }\alpha \leq \phi \leq 2\pi -\alpha \\[1mm] 
0\text{, \ }-\alpha <\phi <\alpha%
\end{array}%
\right.
\end{equation*}%
and that is generated around the $Y-$axis by means of the functions$\
f_{1}\left( x\right) =0$ and $\ f_{2}\left( x\right) =h$. From Eq. (\ref%
{massrevy}) and Eqs.(\ref{revyXa}-\ref{revyZa}) we find%
\begin{equation*}
M=(\pi -\alpha )\rho hb^{2}\ ,
\end{equation*}%
\begin{equation*}
I_{X}=M\left[ \frac{b^{2}}{8}\left( 2-\frac{\sin 2\alpha }{(\pi -\alpha )}%
\right) +\frac{h^{2}}{3}\right] \ ;\ I_{Y}=\frac{Mb^{2}}{2}\ ;\ I_{Z}=M\left[
\frac{b^{2}}{8}\left( 2+\frac{\sin 2\alpha }{(\pi -\alpha )}\right) +\frac{%
h^{2}}{3}\right] \ .
\end{equation*}%
In order to calculate the moments of inertia from axes passing through the
center of mass we use Eqs. (\ref{CMrevyX}-\ref{CMrevyZ}) to get%
\begin{equation*}
X_{CM}=0\text{\ };\text{\ }Y_{CM}=\frac{h}{2}\ ;\ Z_{CM}=-\frac{2}{3}\frac{b%
}{(\pi -\alpha )}\sin \alpha .
\end{equation*}
\end{example}

\begin{example}
\label{bell} MI's for a Gaussian Bell. Let us consider a homogeneous hollow
bell, which can be reasonably described by a couple of Gaussian
distributions (see Fig. \ref{fig:exrevy}).%
\begin{equation*}
f_{1}(x)=Ae^{-\alpha x^{2}}\ ;\ f_{2}(x)=Be^{-\beta x^{2}}\ ,
\end{equation*}%
where $\alpha ,\beta ,A,B$ are positive numbers, $A<B$, and $\alpha >\beta $%
. The MI's are obtained from (\ref{revyYc}, \ref{revyXc})%
\begin{equation*}
I_{Y}=\pi \rho \left[ \frac{B}{\beta ^{2}}-\frac{A}{\alpha ^{2}}\right] \ \
;\ \ I_{X}=I_{Z}=\frac{\pi \rho }{2}\left[ \frac{B}{\beta ^{2}}-\frac{A}{%
\alpha ^{2}}\right] +\frac{\pi \rho }{9}\left[ \frac{B^{3}}{\beta }-\frac{%
A^{3}}{\alpha }\right]
\end{equation*}%
Besides, the mass and the center of mass position read%
\begin{equation*}
M=\pi \rho \left[ \frac{B}{\beta }-\frac{A}{\alpha }\right] \ \ ;\ \ Y_{CM}=%
\frac{1}{4}\left[ \frac{\alpha B^{2}-\beta A^{2}}{\alpha B-\beta A}\right] \
;\ X_{CM}=Z_{CM}=0\ .
\end{equation*}%
When the bell tolls, it rotates around an axis perpendicular to the axis of
symmetry that passes the top of the bell. Thus, this is a real situation in
which the perpendicular MI is required. On the other hand, owing to the
cylindrical symmetry, we can calculate this moment of inertia by taking any
axis parallel to the $X-$axis. In our case the top of the bell corresponds to%
$\ y=B$, and using Steiner's theorem it can be shown that%
\begin{eqnarray*}
I_{X,B} &=&I_{X}+MB(B-2Y_{CM}) \\
I_{X,B} &=&\frac{\pi \rho }{18\alpha ^{2}\beta ^{2}}\left[ \alpha
^{2}B(9+11B^{2}\beta )+\beta ^{2}A(9AB\alpha -2A^{2}\alpha -18B^{2}\alpha -9)%
\right] \ .
\end{eqnarray*}
\end{example}

\subsection{Examples of MI's by the method of contourplots \label%
{ap:contoursexamp}}

\begin{example}
A thin elliptical plate: for this bidimensional object, we can use Eqs. (\ref%
{thinplateX}, \ref{thinplateZ}), the delimited function can be taken from (%
\ref{ellipsedelim}) but with $z=0$. The results are%
\begin{eqnarray*}
I_{X} &=&M\frac{b^{2}}{4}\ \ ;\ \ I_{Y}=M\frac{a^{2}}{4}\ , \\
I_{Z} &=&M\frac{\left( a^{2}+b^{2}\right) }{4}\ \ ;\ \ M=\pi \sigma ab\ .
\end{eqnarray*}%
once again, these axes passes through the center of mass of the figure.
\end{example}

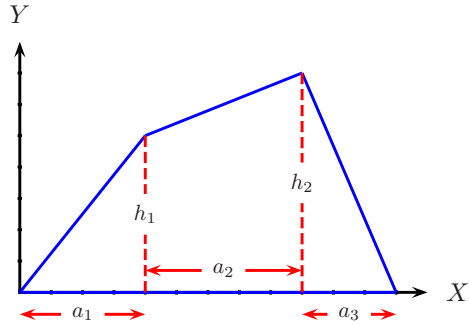
\begin{figure}[!tbh]
\par
\begin{center}
\resizebox{6.0cm}{!}{\mbox{\psset{unit=0.3cm} 
\begin{pspicture}(0,-1)(14,9)
\psaxes[ticksize=0.2pt,labels=none]{->}(13,8)
\psline[linecolor=blue](0,0)(4,5)
\psline[linecolor=blue](4,5)(9,7)
\psline[linecolor=blue](9,7)(12,0)
\psline[linecolor=blue](0,0)(12,0)
\psline[linestyle=dashed,linecolor=red,dash=0.35 0.15](4,0)(4,5)
\psline[linestyle=dashed,linecolor=red,dash=0.35 0.15](9,0)(9,7)
\psline[linecolor=red]{<->}(0,-0.7)(4,-0.7)
\psline[linecolor=red]{<->}(4,0.7)(9,0.7)
\psline[linecolor=red]{<->}(9,-0.7)(12,-0.7)
\rput*[c](4,2.5){\scalebox{0.6}{$h_{1}$}}
\rput*[c](9,3.5){\scalebox{0.6}{$h_{2}$}}
\rput*[c](2,-0.7){\scalebox{0.6}{$a_{1}$}}
\rput*[c](6.5,0.7){\scalebox{0.6}{$a_{2}$}}
\rput*[c](10.5,-0.7){\scalebox{0.6}{$a_{3}$}}
\uput*[u](0,8){\scalebox{0.7}{$Y$}}
\uput*[r](13,0){\scalebox{0.7}{$X$}}

\end{pspicture}
}}
\end{center}
\caption{\textit{Arbitrary quadrilateral, the dimensions are indicated in
the drawing.}}
\label{fig:quadrilat1}
\end{figure}

\begin{example}
An arbitrary quadrilateral, (see Fig. \ref{fig:quadrilat1}): This is a
bidimensional figure, so we apply Eqs. (\ref{thinplateX}, \ref{thinplateZ}).
The bounding functions are given by%
\begin{equation*}
f_{1}\left( x\right) =0\ ,
\end{equation*}%
\begin{equation*}
f_{2}\left( x\right) =\left\{ 
\begin{array}{ccc}
\frac{h_{1}}{a_{1}}x & if & 0\leq x\leq a_{1} \\[2mm] 
\frac{\left( h_{2}-h_{1}\right) }{a_{2}}x+\frac{h_{1}\left(
a_{1}+a_{2}\right) -h_{2}a_{1}}{a_{2}} & if & a_{1}<x\leq a_{1}+a_{2} \\%
[2mm] 
-\frac{h_{2}}{a_{3}}x+\frac{h_{2}}{a_{3}}\left( a_{1}+a_{2}+a_{3}\right) & if
& a_{1}+a_{2}<x\leq a_{1}+a_{2}+a_{3}%
\end{array}%
\right.
\end{equation*}%
and the moments of inertia read%
\begin{equation}
I_{X}=\frac{\sigma }{12}\left[ a_{2}\left( h_{1}+h_{2}\right) \left(
h_{1}^{2}+h_{2}^{2}\right) +h_{1}^{3}a_{1}+h_{2}^{3}a_{3}\right] \allowbreak
\label{quadri1}
\end{equation}%
\begin{eqnarray}
I_{Y} &=&\allowbreak \frac{\sigma }{12}\left[ 12a_{1}a_{2}a_{3}h_{2}+\left(
4a_{1}a_{2}^{2}+3a_{1}^{3}+a_{2}^{3}\right) h_{1}+\left(
4a_{2}a_{3}^{2}+3a_{2}^{3}+a_{3}^{3}\right) h_{2}\right.  \notag \\
&&\left. +6a_{1}^{2}a_{2}\left( h_{1}+h_{2}\right) +4a_{1}h_{2}\left(
2a_{2}^{2}+a_{3}^{2}\right) +6a_{3}h_{2}\left( a_{1}^{2}+a_{2}^{2}\right) 
\right]  \label{quadri2}
\end{eqnarray}%
\begin{equation}
I_{Z}=I_{X}+I_{Y}  \label{quadri3}
\end{equation}%
the center of mass coordinates are given by%
\begin{eqnarray}
x_{CM} &=&\frac{\allowbreak \sigma }{6M}\left[ 3a_{1}a_{2}\left(
h_{1}+h_{2}\right) +3a_{3}h_{2}\left( a_{1}+a_{2}\right) \right.  \notag \\
&&\left. +h_{1}\left( 2a_{1}^{2}+a_{2}^{2}\right) +h_{2}\left(
2a_{2}^{2}+a_{3}^{2}\right) \right]  \label{quadri4}
\end{eqnarray}%
\begin{equation}
y_{CM}=\frac{\sigma }{6M}\left[ a_{2}h_{1}h_{2}+\left( a_{1}+a_{2}\right)
h_{1}^{2}+\left( a_{2}+a_{3}\right) h_{2}^{2}\right] \allowbreak
\label{quadri5}
\end{equation}%
with%
\begin{equation}
M=\frac{\sigma }{2}\left[ a_{1}h_{1}+a_{2}\left( h_{1}+h_{2}\right)
+a_{3}h_{2}\right]  \label{quadri6}
\end{equation}%
The quantities $I_{X},~y_{CM},$ and$\ M$; are invariant under traslations in 
$x$, for example $I_{X}$ might be calculated as%
\begin{equation*}
I_{X}=\frac{\sigma }{3}\int_{x_{0}}^{x_{f}}f_{2}\left( x\right) ^{3}dx=\frac{%
\sigma }{3}\int_{x_{0}-\Delta x}^{x_{f}-\Delta x}f_{2}\left( u+\Delta
x\right) ^{3}du
\end{equation*}%
where we have performed a traslation $\Delta x$ to the left. It is
equivalent to make the change of variables $u=x-\Delta x$. This property can
be used to evaluate the integrals easier. Specifically, the piece of Fig. %
\ref{fig:quadrilat1} lying in the interval $a_{1}<x\leq a_{1}+a_{2}$, can be
traslated to the origin by using $\Delta x=a_{1}$; and the piece of this
figure lying at $a_{1}+a_{2}<x\leq a_{1}+a_{2}+a_{3}$ can be also traslated
to the origin with $\Delta x=a_{1}+a_{2}$. On the other hand, though the
quantities $I_{Y}$ and $x_{CM}$ are not invariant under such traslations,
the same change of variables simplifies their calculations. This strategy is
very useful in solids or surfaces that can be decomposed by pieces (i.e.
when at least one of the generator functions is defined by pieces). For
example, the same change of variables could be used if we are interested in
the solid of revolution generated by the surface of Fig. \ref{fig:quadrilat1}%
.

Finally, from Eqs. (\ref{quadri1}-\ref{quadri6}) we can obtain many
particular cases, some of them are

\begin{itemize}
\item $h_{1}=h_{2}$ trapezoid.

\item $a_{1}=h_{1}=0$ arbitrary triangle.

\item $a_{2}=0,\ a_{1}=a_{3}=\frac{h_{1}}{\sqrt{3}}=\frac{h_{2}}{\sqrt{3}}%
\equiv \frac{L}{2}$ equilateral triangle of side length $L$.

\item $a_{1}=h_{1}=0,\ a_{2}=a_{3}=\frac{h_{2}}{\sqrt{3}}\equiv \frac{L}{2}$%
\ equilateral triangle of side length $L$.

\item $a_{1}=a_{3}=h_{1}=0$ triangle with a right angle.

\item $h_{2}=h_{1},$ $a_{2}=0$ arbitrary triangle.

\item $a_{1}=a_{3}=0,\ h_{2}=h_{1}$ rectangle.
\end{itemize}
\end{example}

\section{Table of moments of inertia\label{ap:tables}}

\begin{figure}[tbh]
\begin{center}
\includegraphics[width=14cm]{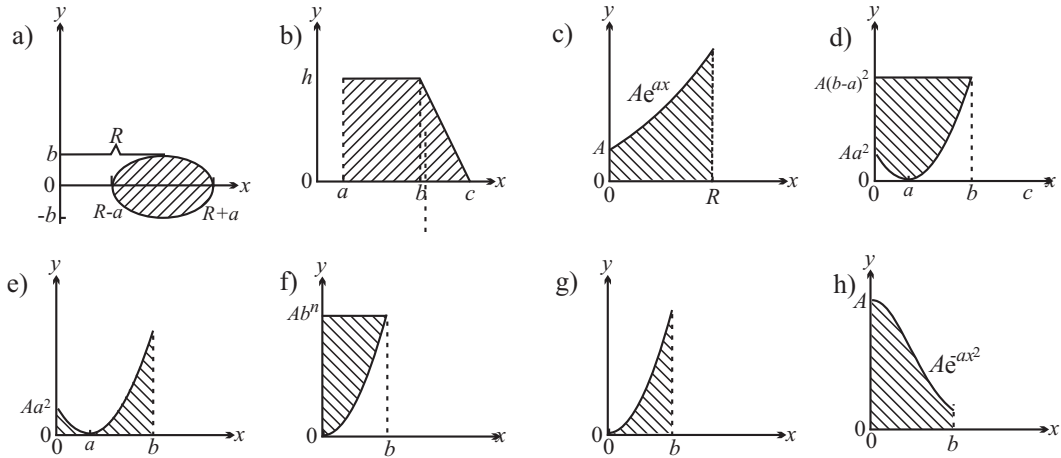}
\end{center}
\caption{\textit{Surfaces that generates the solids whose moments of inertia
appears on the table \protect\ref{tab:gentab}}}
\label{fig:figtab}
\end{figure}
In table \ref{tab:gentab} on page \pageref{tab:gentab}, the moments of
inertia for a variety of solids of revolution generated around the $Y-$axis
are displayed, such table includes the function generators and any other
information necessary to carry out the calculations by means of our methods.
The surfaces that generates the solids are displayed in Fig. \ref{fig:figtab}%
. Observe that the first of these surfaces generates a torus with elliptical
cross section and the second one generates a truncated hollow cone.

Finally, there are some conditions for certain parameters of these figures.
In Fig. (d), $a>0$, and $b>2a$; in Fig. (e) $b>0$, and $a>0$, in Fig. (f), $%
n>0$; for Fig. (g), $n>-2/3$; in Figs. (c) and (h) $a$ can also be
negative.\newpage

\begin{table}[tbh]
\begin{center}
\scalebox{0.93}[0.83]{\rotatebox{90}{\mbox{\renewcommand{\arraystretch}{1.7}
\begin{tabular}{|c|c|c|c|c|c|c|c|c|}
\hline
Fig & $f_{1}(x)$ & $f_{2}(x)$ & $x_{0}$ & $x_{f}$ & $M$ & $I_{Y}$ & $I_{X}=I_{Z}$ & $Y_{CM}$ \\[2mm]\hline
a & $-f_{2}(x)$ & $\frac{b\sqrt{a^{2}-(x-R)^{2}}}{a}$ & $R-a$ & $R+a$ & $2\pi ^{2}\rho Rba$ & $M(R^{2}+\frac{3}{4}a^{2})$ & $\frac{M}{8}(4R^{2}+3a^{2}+2b^{2})$ & $0$ \\[2mm]\hline
b & $0$ & 
\begin{tabular}{l}
$h,a<x<b$ \\ 
$h\frac{c-x}{c-b},b\leq x\leq c$\end{tabular}$\ $ & $a$ & $c$ & $\begin{tabular}{l}
$\frac{\pi \rho h}{3}(b^{2}+bc$ \\ 
$+c^{2}-3a^{2})$\end{tabular}$ & 
\begin{tabular}{l}
$\frac{\pi \rho h}{10}[b^{4}+b^{3}c+b^{2}c^{2}$ \\ 
$+bc^{3}+c^{4}-5a^{4}]$\end{tabular}
& $\begin{tabular}{l}
$\frac{I_{y}}{2}+\pi \rho h^{3}\times $ \\ 
$\frac{\left( 3bc+6b^{2}+c^{2}-10a^{2}\right) }{30}$\end{tabular}$ & $\frac{h}{4}\frac{\left( 3b^{2}+2bc+c^{2}-6a^{2}\right) }{[b^{2}+bc+c^{2}-3a^{2}]}$ \\[2mm]\hline
c & $0$ & $Ae^{ax}$ & $0$ & $b$ & $\begin{tabular}{l}
$\frac{2\pi A\rho }{a^{2}}\left( e^{ab}ab\right. $ \\ 
$\left. -e^{ab}+1\right) $\end{tabular}$ & $\begin{tabular}{l}
$\frac{2\pi A\rho }{a^{4}}e^{ab}\left( a^{3}b^{3}-3a^{2}b^{2}\right. $ \\ 
$\left. +6\left( ab-1+e^{-ab}\right) \right) $\end{tabular}$ & $\begin{tabular}{l}
$\frac{1}{2}I_{Y}+\frac{2\pi }{27a^{2}}\rho A^{3}\times $ \\ 
$\left( e^{3ab}\left( 3ab-1\right) +1\right) $\end{tabular}$ & $\frac{A}{8}\frac{\left( e^{2ab}\left( 2ab-1\right) +1\right) }{\left(
e^{ab}\left( ab-1\right) +1\right) }$ \\[2mm]\hline
d & $A(x-a)^{2}$ & $A(b-a)^{2}$ & $0$ & $b$ & $\allowbreak \frac{\pi A\rho
b^{3}}{6}\left( 3b-4a\right) $ & $M\frac{b^{2}}{5}(\frac{5b-6a}{3b-4a})$ & $\begin{tabular}{l}
$\frac{I_{y}}{2}+\frac{b^{3}\rho A^{3}\pi }{84}(21b^{5}$ \\ 
$-120ab^{4}+280a^{2}b^{3}$ \\ 
$-336a^{3}b^{2}+210a^{4}b-56a^{5})$\end{tabular}$\  & $\frac{A\left( 10b^{3}+45a^{2}b-36ab^{2}-20a^{3}\right) }{5\left(
3b-4a\right) }$ \\[2mm]\hline
e & $0$ & $A(x-a)^{2}$ & $0$ & $b$ & $\begin{tabular}{l}
$\frac{1}{6}\pi \rho Ab^{2}(3b^{2}$ \\ 
$-8ab+6a^{2})$\end{tabular}$ & $\begin{tabular}{l}
$\frac{\pi \rho b^{4}A}{30}(10b^{2}$ \\ 
$-24ab+15a^{2})$\end{tabular}$ & $\begin{tabular}{l}
$\frac{I_{y}}{2}+\frac{b^{2}\rho A^{3}\pi }{84}\left[ 28a^{6}+7b^{6}-\right. 
$ \\ 
$48ab^{5}-112a^{5}b+140a^{2}b^{4}$ \\ 
$\left. -224a^{3}b^{3}+210a^{4}b^{2}\right] $\end{tabular}\ $ & $\begin{tabular}{l}
$\frac{A}{5}(15a^{4}+5b^{4}-24ab^{3}$ \\ 
$-40a^{3}b+45a^{2}b^{2})\times $ \\ 
$\frac{1}{\left( 3b^{2}-8ab+6a^{2}\right) }$\end{tabular}$ \\[2mm]\hline
f & $Ax^{n}$ & $Ab^{n}$ & $0$ & $b$ & $\pi \rho Ab^{n+2}\frac{n}{n+2}$ & $M\frac{b^{2}}{2}\frac{n+2}{n+4}$ & $\frac{I_{y}}{2}+MA^{2}b^{2n}\frac{n+2}{3n+2}$ & $A\frac{n+2}{2n+2}b^{n}$ \\[2mm]\hline
g & $0$ & $Ax^{n}$ & $0$ & $b$ & $2\pi \rho A\frac{b^{n+2}}{n+2}$ & $Mb^{2}\frac{n+2}{n+4}$ & $\frac{I_{y}}{2}+\frac{1}{3}M\frac{n+2}{3n+2}A^{2}b^{2n}$
& $\frac{A}{4}\frac{b^{n}(n+2)}{n+1}$ \\[2mm]\hline
h & $0$ & $Ae^{-ax^{2}}$ & $0$ & $b$ & $\pi \rho A\frac{1-e^{-b^{2}a}}{a}$ & 
$M\frac{1-e^{-ab^{2}}(1-ab^{2})}{a(1-e^{-ab^{2}})}$ & $\frac{I_{Y}}{2}+\frac{MA^{2}(1+e^{-ab^{2}}+e^{-2ab^{2}})}{9}$ & $\frac{A}{4}\frac{(1-e^{-2ab^{2}})}{(1-e^{-ab^{2}})}$ \\[2mm]\hline
\end{tabular}}}}
\end{center}
\caption{Moments of inertia for a variety of solids of revolution generated
around the $Y-$ axis, the surfaces that generates the solids are displayed
in Fig. \protect\ref{fig:figtab}}
\label{tab:gentab}
\end{table}


\begin{thebibliography}{9}
\bibitem{mecanica} D. Kleppner and R. Kolenkow, \emph{An introduction to
mechanics} (McGRAW-HILL KOGAKUSHA LTD, 1973); R. Resnick and D. Halliday, 
\emph{Physics} (Wiley, New York, 1977), 3rd Ed.; M. Alonso and E. Finn, 
\emph{Fundamental University Physics, Vol I, Mechanics} (Addison-Wesley
Publishing Co., Massachussets, 1967).

\bibitem{engineering} R. C. Hibbeler, \emph{Engineering Mechanics Statics},
Seventh Ed. (Prentice-Hall Inc., New York,1995).

\bibitem{calculo} Louis Leithold, \emph{The Calculus with Analytic Geometry}
(Harper \& Row, Publishers, Inc. , New York, 1972), Second Ed.; E. W.
Swokowski, \emph{Calculus with Analytic Geometry} (PWS-KENT Publishing Co.,
Boston Massachusetts, 1988), Fourth Ed.; S. K. Stein, \emph{Calculus and
Analytic Geometry} (Mc-Graw Hill Book Co. 1987), Fourth Ed.

\bibitem{methods} R. Szmytkowski, \textquotedblleft Simple method of
calculation of moments of inertia\textquotedblright , Am. J. Phys. \textbf{56%
}, 754-756 (1988); R. Rabinoff \textquotedblleft Moments of inertia by
scaling arguments: How to avoid messy integrals\textquotedblright\ Am. J.
Phys. \textbf{53}, 501-502 (1985).

\bibitem{mathprop} Carl M. Bender and Lawrence R. Mead, \textquotedblleft
D-dimensional moments of inertia\textquotedblright\ Am. J. Phys. \textbf{63}%
, 1011-1014 (1995); J. Casey and S. Krishnaswamy, \textquotedblleft Problem:
Which rigid bodies have constant inertia tensors?\textquotedblright\ Am. J.
Phys. \textbf{63}, 276-281 (1995); P. K. Aravind, \textquotedblleft A
comment on the moment of inertia of symmetrical solids\textquotedblright\
Am. J. Phys. \textbf{60}, 754-755 (1992); P. K. Aravind, \textquotedblleft
Gravitational collapse and moment of inertia of regular polyhedral
configurations\textquotedblright\ Am. J. Phys. \textbf{59}, 647-652 (1991);
J. Satterly, \textquotedblleft Moments of Inertia of Solid Rectangular
Parallelopipeds, Cubes, and Twin Cubes, and Two Other Regular
Polyhedra\textquotedblright\ Am. J. Phys. \textbf{25}, 70-78 (1957) ; J.
Satterly, \textquotedblleft Moments of Inertia of Plane
Triangles\textquotedblright\ Am. J. Phys. \textbf{26}, 452-453 (1958).

\bibitem{exp} W. N. Mei and Dan Wilkins, \textquotedblleft Making a pitch
for the center of mass and the moment of inertia\textquotedblright\ Am. J.
Phys. \textbf{65}, 903-907 (1997); Joseph C. Amato and Roger E. Williams and
Hugh Helm, \textquotedblleft A \emph{black box }moment of inertia
apparatus\textquotedblright\ Am. J. Phys. \textbf{63}, 891-894 (1995).

\bibitem{theorem} J. F. Streib, \textquotedblleft A theorem on moments of
inertia\textquotedblright\ Am. J. Phys. \textbf{57}, 181 (1989).

\bibitem{revus} Rodolfo A. Diaz, William J. Herrera, R. Martinez,
http://xxx.lanl.gov/list/physics/0507 \ Preprint number: physics/0507172.

\bibitem{variac1} G. Arfken. \emph{Mathematical methods for physicists},
Second Ed. (Academic Press, International Edition, 1970) Chap. 17.
\end{thebibliography}
\end{document}